\documentclass[
	aps,
    prd,
	reprint,
	amssymb,amsmath,
	nofootinbib,
	superscriptaddress,
	notitlepage,
]{revtex4-1}

\usepackage{graphicx}
\usepackage{dcolumn}
\usepackage{bm}
\usepackage{color}
\usepackage{xspace}
\usepackage{ulem}
\usepackage{multirow}
\def\apjl{Astrophys.\ J.\ Lett.\ }

\def\mnras{Mon.\ Not.\ Roy.\ Astron.\ Soc.\ }

\newcommand{\no}{{\dot n}_0}

\newcommand{\betaz}{\beta_{z}}

\newcommand{\Msol}{M_{\odot}}

\begin{document}

\title{Dark Matter Spike surrounding Supermassive Black Holes Binary and the Nanohertz Stochastic Gravitational Wave Background}

\author{Zhao-Qiang Shen}\email{zqshen@pmo.ac.cn}
\affiliation{Key Laboratory of Dark Matter and Space Astronomy, Purple Mountain Observatory, Chinese Academy of Sciences, Nanjing 210023, China}

\author{Guan-Wen Yuan}
\email{guanwen.yuan@unitn.it}
\affiliation{Department of Physics, University of Trento, Via Sommarive 14, 38123 Povo (TN), Italy}
\affiliation{Trento Institute for Fundamental Physics and Applications (TIFPA)-INFN, Via Sommarive 14, 38123 Povo (TN), Italy}

\author{Yi-Ying Wang}\email{wangyy@pmo.ac.cn}
\affiliation{Key Laboratory of Dark Matter and Space Astronomy, Purple Mountain Observatory, Chinese Academy of Sciences, Nanjing 210023, China}

\author{Yuan-Zhu Wang}\email{vamdrew@zjut.edu.cn}
\affiliation{Institute for Theoretical Physics and Cosmology, Zhejiang University of Technology, Hangzhou, 310032, People’s Republic of China}

\author{Yin-Jie Li}\email{liyinjie@pmo.ac.cn}
\affiliation{Key Laboratory of Dark Matter and Space Astronomy, Purple Mountain Observatory, Chinese Academy of Sciences, Nanjing 210023, China}

\author{Yi-Zhong Fan}
\email{yzfan@pmo.ac.cn}
\affiliation{Key Laboratory of Dark Matter and Space Astronomy, Purple Mountain Observatory, Chinese Academy of Sciences, Nanjing 210023, China}
\affiliation{School of Astronomy and Space Science, University of Science and Technology of China, Hefei 230026, China}

\begin{abstract}

The NANOGrav, PPTA, EPTA, CPTA and MPTA collaborations have reported compelling evidence for the existence of the Stochastic Gravitational-Wave Background (SGWB). This inferred background's amplitude and frequency spectrum align closely with the astrophysical predictions for a signal originating from the population of supermassive black hole (SMBH) binaries.
Considering these findings, we explore the possibility of detecting dark matter (DM) spikes surrounding SMBHs, which could alter the gravitational-wave waveform and influence the SGWB. We show that the evolution of SMBH binaries, driven by both gravitational radiation and the dynamic friction of the surrounding DM spike, presents observable effects in the nHz frequency domain of the SGWB.
We also employ the Bayesian inference method to fit the SGWB spectra from the NANOGrav, EPTA, and PPTA.
The model with DM spike improves the fittings to the former two data sets.
The spike slope $\gamma_{\rm sp}$ is slightly smaller than 1, which may suggest that the spike is flattened during the inspiral of the SMBHBs.

\end{abstract}

\date{\today}

\maketitle
 
\section{Introduction}\label{sec:introduction}

Gravitational waves (GWs) were initially predicted by Einstein in General Relativity, but their actual detection through the orbital decay of a binary pulsar system PSR B1913+16 \citep{Weisberg:2004hi} took more than half a century. And almost a century had to elapse until their direct measurement (GW150914) from binaries of stellar-mass black holes \citep{LIGOScientific:2016aoc}. 
In most galaxies, supermassive black holes (SMBHs) with masses of $10^5 - 10^{10} M_{\odot}$ reside at the galactic centers.
These SMBHs are expected to form close binary systems long after their host galaxies have merged.
Supermassive black hole binaries (SMBHBs)—broadly defined by masses of $10^8-10^{10} M_{\odot}$ emit GWs with slowly evolving frequencies, contributing to a noise-like broadband signal (the GW background, GWB) in the nHz range.
Pulsar timing array (PTA) experiments, such as the North American Nanohertz Observatory for Gravitational Waves (NANOGrav)~\cite{NANOGrav:2020bcs}, European Pulsar Timing Array (EPTA) \cite{Chen:2021rqp}, Parkes Pulsar Timing Array (PPTA)~\cite{Goncharov:2021oub} have reported a common spectrum noise at frequencies around $10^{-8}$ Hz.

PTA collaborations including
NANOGrav~\cite{NANOGrav:2023gor},
EPTA~\cite{Antoniadis:2023ott}, PPTA~\cite{Reardon:2023gzh},
the Chinese PTA~\cite{Xu:2023wog} and
the MeerKAT PTA~\cite{2025MNRAS.536.1467M}
have jointly announced compelling evidence supporting the existence of a stochastic gravitational wave background (SGWB) within the frequency range of approximately $1 - 10 ~\mathrm{nHz}$,
which attracts lots of implications (e.g the reviews~\cite{NANOGrav:2023hvm, Antoniadis:2023xlr}).
Earlier hints of such a background appeared in the NANOGrav 12.5 years dataset, sparking scientific interest and prompting discussions on the potential origin of this signal within the nano-Hertz frequency range. Among the most widely discussed explanations is the population of inspiraling SMBHBs, which are expected to be the dominant astrophysical source of the SGWB in this frequency band~\cite{Burke-Spolaor:2018bvk, Sesana:2008mz,Broadhurst:2023tus}.
Assuming the evolution of these binaries is governed solely by the emission of gravitational radiation from quasi-circular orbits, the resulting GWB follows a characteristic strain spectrum scaling as power-law $f^{-2/3}$~\cite{Sesana:2008mz,Burke-Spolaor:2018bvk, NANOGrav:2023hfp, NANOGrav:2023pdq}.

Although current observations are consistent with the SGWB originating from inspiraling SMBHBs, a variety of cosmological processes -- beyond the canonical astrophysical scenario -- may also contribute to the signal within the nanohertz frequency band~\cite{Franciolini:2023wjm, Lambiase:2023pxd, Han:2023olf, Guo:2023hyp, Wang:2023len, Ellis:2023tsl, Vagnozzi:2023lwo, Liu:2023ymk, Wu:2023hsa,Fujikura:2023lkn, Kitajima:2023cek, Franciolini:2023pbf, Megias:2023kiy, Ellis:2023dgf, Bai:2023cqj, Yang:2023aak, Ghoshal:2023fhh, Deng:2023btv, Mitridate:2023oar, King:2023cgv, Zu:2023olm, Li:2023yaj, Addazi:2023jvg, Liu:2023ymk, Konoplya:2023fmh, Unal:2023srk,Broadhurst:2023tus,Bian:2023dnv,Madge:2023cak,Balaji:2023ehk,Liu:2023pau,Vagnozzi:2023lwo,Madge:2023dxc,Ellis:2023oxs, Liang:2023pbj,Calza:2024qxn}. 
One well-motivated example involves enhanced scalar perturbations at small scales during inflation, which may lead to the formation of primordial black holes (PBHs)~\cite{Ivanov:1994pa, Carr:2016drx, Carr:2020xqk, Chen:2019xse, Yuan:2023bvh,Liu:2023ymk,Wang:2023ost,Inomata:2023zup,Franciolini:2023pbf,Figueroa:2023zhu}. These scalar perturbations can also source second-order scalar-induced gravitational waves, potentially contributing to the SGWB~\cite{Boyle:2005se, Ananda:2006af,Baumann:2007zm, Yuan:2021qgz,Saito:2009jt,Wang:2019kaf,Zhao:2022kvz,Papanikolaou:2020qtd,Domenech:2021ztg,Balaji:2023ehk,Cai:2023dls,Domenech:2024rks,Cecchini:2025oks}. Other prominent cosmological sources include first-order phase transitions in the early universe, as well as the dynamics of topological defects such as cosmic strings and domain walls~\cite{Kosowsky:1992rz, Kamionkowski:1993fg,Caprini:2007xq,Hindmarsh:2013xza,Kibble:1976sj,Vilenkin:1981bx,Hogan:1984is,Caldwell:1991jj,Vilenkin:1981zs,Chang:1998tb,Hiramatsu:2010yz,Chen:2022azo,Ashoorioon:2022raz,Bian:2022qbh,Athron:2023xlk,He:2023ado}. These exotic mechanisms --ranging from cosmological to physical processes-- are reviewed in~\cite{Burke-Spolaor:2018bvk,NANOGrav:2023hvm, Antoniadis:2023xlr}.  The SGWB from such cosmological sources offers a powerful window into physics beyond the Standard Model and provides unique insight into the conditions of the early universe.

The nature of dark matter (DM), an invisible component that provides the additional gravitational force necessary to explain a range of phenomena across different scales, remains one of the most significant enigmas in the universe~\cite{Bertone:2004pz, Clowe:2006eq}. In scenarios involving massive DM particles, their interaction with SMBHs can lead to the formation of steep density enhancements known as dark matter spike, where the DM profile becomes even steeper than the classical cusp~\citep{Ullio:2001fb, Yuan:2021mzi, Shen:2023kkm}. 
A variety of theoretical studies have explored the properties and potential signatures of such spike~\cite{Broadhurst:2023tus,Ghoshal:2023fhh,Aghaie:2023lan,Hu:2023oiu, Dosopoulou:2023umg, Zhang:2025mdl, Mitra:2025tag}. Numerous observational strategies have also been proposed to identify these features, including gravitational wave detection~\cite{Eda:2014kra, Yue:2019ozq, Li:2021pxf,Zhang:2024hrq, Cheng:2024mgl, Daniel:2025mna,Zhang:2025mdl,Xie:2025udx,Feng:2025fkc}, analysis of stellar orbital dynamics~\cite{Shen:2023kkm}, searching for DM annihilation signatures that depend on the spike profile~\cite{Alonso-Alvarez:2024gdz,Chen:2025jch}, and studies of DM-induced dynamical friction~\cite{2024ApJ...962L..40C,Fischer:2024dte}, among others.

In this work, we investigate the impact of DM spikes surrounding the SMBHBs on modifying the nanohertz frequency SGWB. Specifically, we show that the presence of such spikes can imprint detectable signatures in the SGWB by modulating the amplitude of GW emission from inspiraling SMBHBs. It is worth noting that astrophysical uncertainties-such as galaxy pairs and gas/stellar distribution in the center of SMBH-can influence the predicted SGWB~\cite{Sesana:2006xw, Sesana:2006ne, Sesana:2007vr, Sesana:2008xk, Shannon:2015ect, Kelley:2016gse}. While a detailed treatment of these uncertainties is beyond the scope of this work, it will be addressed in future studies. Our current focus is to assess the feasibility of identifying DM spikes through their imprint on the SGWB.

The structure of the paper is outlined as follows: In Section~\ref{sec:dynafric and spike}, we begin by discussing the concept of dynamical friction arising from the distribution of dark matter spikes surrounding SMBHBs. Section~\ref{sec:SGWB} provides a concise overview of the methodology employed to derive the spectrum of the SGWB, which involves utilizing the equation of motion for the SMBHs in conjunction with their energy balance equation. We proceed by detailing our data analysis and presenting our findings in Section~\ref{sec:data_analysis}. Lastly, we conclude by providing a summary of our conclusions in Section~\ref{sec:conclusion}.

\section{Dynamical Friction induced by Dark Matter Particles}\label{sec:dynafric and spike}
In the vicinity of a supermassive black hole (SMBH), the density of the surrounding dark matter (DM) can increase substantially, particularly if there is a DM spike present, as we have previously addressed. Within these settings, the gravitational interaction between the SMBH and DM particles results in dynamical friction. This is a drag force that acts to decelerate the SMBH by reducing its kinetic energy and angular momentum. Initially formulated by Chandrasekhar for stellar systems~\cite{1943ApJ....97..255C}, this principle has been extensively adapted to explore the influence of dark matter on black hole dynamics and mergers.

\subsection{The NFW Spike Distribution}\label{sec:spike}
The most popular DM density model is the NFW profile~\citep{Navarro:1996gj}. The density at the Galactocentric radius $r$ is
\begin{equation}\label{eq::gnfw_halo}
    \rho_{\rm nfw}(r)=\frac{\rho_0}{(r/r_{\rm s}) (1+r/r_{\rm s})^2}.
\end{equation}
Here, $r_{\rm s}$ is the scale radius with the condition ${\rm d}  {\rm ln}\rho/{\rm d} {\rm ln} r=-2$ and $\rho_0$ is the density normalization.
The $r_{\rm s}$ and $\rho_0$ of a galaxy at the redshift $z$ are associated with the DM halo mass $M_{200}$ through
\begin{equation}
    M_{200} \equiv 200\times \frac{4}{3}\pi\rho_{\rm c}(z) R_{200}^3
       = \int_0^{R_{200}} 4\pi r^2 {\rm d}r \rho_{\rm nfw}(r).
\end{equation}
$\rho_{\rm c}(z)$ is the critical density at redshift $z$: 
$\rho_{\rm c}(z) = 3H(z)^2/(8\pi G)$,
where $H(z)=H_0/\sqrt{\Omega_\Lambda + \Omega_m (1+z)^3}$ and $H_0 \equiv 100h~{\rm km\,s^{-1}}=70~\rm km\,s^{-1}\,Mpc^{-1}$, $\Omega_\Lambda=0.6847$, $\Omega_{\rm m}=1-\Omega_\Lambda$~\citep{Planck:2018vyg}, and $G$ is the gravitational constant..
The size of the dark matter $R_{200}$ is related to the concentration $c_{\rm 200}(M_{200},z)\equiv R_{\rm 200}/r_{\rm s}$.
The concentration can be calculated with~\citep{Klypin:2014kpa}
\begin{equation}
    c_{200} = C_{\rm c}(z) \left( \frac{M_{200}}{10^{12}h^{-1}M_\odot} \right)^{-\gamma_{\rm c}(z)} \left[ 1+\left(\frac{M_{200}}{M_{\rm c0}(z)}\right)^{0.4} \right],
\end{equation}
where the parameters $C_{\rm c}(z)$, $\gamma_{\rm c}(z)$ and $M_{\rm c0}(z)$ are the interpolate functions given the results in Tab.2 of~\citep{Klypin:2014kpa}.

We further relate the halo mass to the mass of the central BH.
The halo mass is firstly converted to the stellar mass through the relation~\citep{Girelli:2020goz}
\begin{equation}
    \frac{M_{\rm star}}{M_{200}}(z) = \frac{2A_{\rm sh}(z)}{\left[ \left(\frac{M_{200}}{M_{\rm sh}(z)}\right)^{-\beta_{\rm sh}(z)} + \left(\frac{M_{200}}{M_{\rm sh}(z)}\right)^{\gamma_{\rm sh}(z)} \right]},
\end{equation}
where $\lg M_{\rm sh}(z)=B_{\rm sh} + z\cdot \mu_{\rm sh}$, $A_{\rm sh}(z) = C_{\rm sh}\cdot(1+z)^{\nu_{\rm sh}}$, $\gamma_{\rm sh}(z)=D_{\rm sh}\cdot(1+z)^{\eta_{\rm sh}}$, and $\beta_{\rm sh}(z) = F_{\rm sh}\cdot z+E_{\rm sh}$.
We use the best-fit parameter set in Tab.3 of~\citep{Girelli:2020goz}:
$B_{\rm sh}=11.79$, $C_{\rm sh}=0.046$, $D_{\rm sh}=0.709$, $E_{\rm sh}=0.96$, $F_{\rm sh}=0.043$, $\mu_{\rm sh}=0.20$, $\nu_{\rm sh}=-0.38$, and $\eta_{\rm sh}=-0.18$.
The stellar mass of a galaxy is approximately related to the bulge mass through~\citep{Chen:2018znx}
\begin{equation}
    \frac{M_{\rm bul}}{M_{\rm star}}= 
        \begin{cases}
            0.615  & \quad M_{\rm star,10}\leq1 \\
            0.615 + \frac{\sqrt{6.9}}{(\lg M_{\rm star,10} )^{1.5}}e^{-{\frac{3.45}{\lg M_{\rm star,10}}}} & \quad M_{\rm star,10}>1,
        \end{cases}
\end{equation}
where $M_{\rm star,10}\equiv M_{\rm star}/10^{10}M_\odot$.
According to~\citep{Kormendy:2013dxa}, there is a relationship between the bulge mass and the mass of the central black hole, represented by the equation $M_{\bullet}=4.9\times10^8 M_\odot (M_{\rm bul}/10^{11}M_\odot)^{1.17}$. 
We consider that the ultimate mass of the SMBH in the galaxy is equivalent to the combined mass of the two black holes, $M_{\rm tot}$. Consequently, the associated Navarro-Frenk-White (NFW) profile can be derived from this assumption.

\begin{figure}
    \centering
    \includegraphics[width=0.95\linewidth]{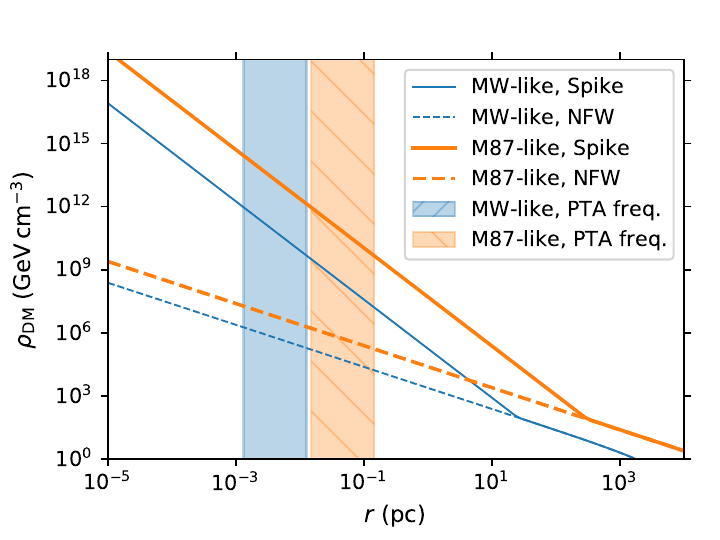}
    \caption{The predicted DM density distributions inside the MW-like (narrow blue lines) and M87-like (wide orange lines) galaxies.
    The total mass of the SMBH binary in MW-like (M87-like) galaxy is assumed to be $4\times10^6M_\odot$ ($6\times10^9M_\odot$).
    The solid and dashed lines are used to represent the NFW profile and NFW+spike scenario, respectively.
    The color bands correspond to the PTA frequency band between $0.03~\rm yr^{-1}$ and $0.9~\rm yr^{-1}$.}
    \label{fig:spike}
\end{figure}

For the DM distribution with a spike, we adopt the piece-wise function~\cite{Lacroix:2018zmg}
\begin{equation}\label{eq::nfw_combine}
    \rho_{\rm sp}(r) =
    \begin{cases}
        0 & \qquad r \leq 2R_{\rm sch} , \\
        \rho_{\rm sp0} \left( \frac{r}{R_{\rm sp}}\right)^{-\gamma_{\rm sp}} & \qquad 2R_{\rm sch} < r \leq R_{\rm sp}, \\
        \rho_{\rm nfw}(r) & \qquad r > R_{\rm sp}.
    \end{cases}
\end{equation}
where $R_{\rm sp}=a_\gamma r_{\rm s} \sqrt{M_{\rm tot}/(\rho_0 r_{\rm s}^3)}$ is the spike radius, the $a_\gamma$ is the scale factor which is approximately $0.1$ for the NFW spike~\cite{Gondolo:1999ef}. $\gamma_{\rm sp}$ is the slope of the DM spike and is $\gamma_{\rm sp,ad}=7/3$ in the environment when the BH grows adiabatically. Considering the DM may be flattened during the inspiring of the BHs (e.g.~\citep{Ullio:2001fb}), we set $\gamma_{\rm sp}$ as a free parameter in the fitting. 

Fig.~\ref{fig:spike} shows the DM density profile inside the Milky Way (MW)-like and M87-like galaxies, drawn in narrow blue lines and wide orange lines respectively.
The solid and dashed lines correspond to the NFW profile and the NFW+spike profile.
The color bands illustrates the radius region detectable by the PTA experiments.

\subsection{Dynamical Friction of Dark Matter Particles}\label{sec:dynafric}

In the vicinity of a SMBH, the density of the surrounding DM can increase substantially, such as the DM spike we have previously addressed. Within this settings, the gravitational interaction between the SMBH and DM particles results in dynamical friction. This is a drag force that acts to decelerate the SMBH by reducing its kinetic energy and angular momentum. Initially formulated by Chandrasekhar for stellar systems~\cite{1943ApJ....97..255C}, this mechanism has been extensively adapted to explore the influence of DM on black hole dynamics and mergers.

The process arises as a massive object of mass $M$ and velocity $\textbf{v}$ travels through a background of lighter particles (mass $m$, number density $\rho$), generatign a gravitational wake. The overdense wake exerts a net pull on the object, resulting in a decelerating force:
\begin{equation}
f_{\text{df}} = 4\pi G^2 M^2 \rho_{\text{DM}}(r) \ln \Lambda /v^2, 
\end{equation}
where $\rho_{\rm DM}(r)$ is the local DM density and ${\rm ln}\Lambda$ is the Coulomb logarithm, approximated by
$\Lambda \approx b_{\text{max}} v_{\text{typ}}^2 / \left(G\mu\right)$, with the reduced mass $\mu=\frac{M_1 M_2}{M_1 + M_2}$ and the maximum impact parameter $b_{\rm max}$ . In dense DM environment, such as spikes near SMBHs, this friction can notably affect orbital evolution on astrophysical timescales.

\section{Modeling the SGWB spectrum}\label{sec:SGWB}
In this section, we follow the formalism developed in Ref~\cite{Eda:2014kra} to derive the spectrum of the SGWB generated by a population of inspiraling SMBH binaries. Our approach combines the equations of motion with energy balance considerations, incorporating both GW emission and dynamical friction effect.

\subsection{Equation of Motion for the SMBH Binary}\label{sec:EoM_SMBHB}
We consider a SMBHB, which involves two massive compact objects with masses of $M_1$ and $M_2$, respectively. By adopting a reference frame attached to the barycenter, the equations of motion of the radial relative separation between the two black holes can be approximated with
\begin{equation}
\begin{aligned}
 \dfrac{{\rm d}^2 r_1}{{\rm d}t^2} &= - \dfrac{GM_2}{R^2} + \omega_s^2 r_1 , \\
 \dfrac{{\rm d}^2 r_2}{{\rm d}t^2} &= - \dfrac{GM_1}{R^2} + \omega_s^2 r_2 ,
 \label{Eq:EOM} 
\end{aligned}
\end{equation}
where $R=r_1 + r_2$ is the distance between two SMBHs, $r_1$ and $r_2$ are the radii of the $M_1$ and $M_2$'s orbit around the barycenter, and $\omega_s$ is the angular velocity of the binary system.
Here the gravitational force contributed by the extended DM mass and the GW back-reaction force are neglected because these effects are much smaller than the gravitational potential of the SMBH in a single orbit.
We will introduce these effects to include an adiabatic evolution of the orbital radius in the next subsection.

We assume that these two SMBHs orbit in a circular manner for simplicity, 
{since GW emission always causes the eccentricity to decrease}. The orbital radius $R$ is obtained by solving ${\rm d}^2r/{\rm d}t^2=0$ in Eq. (\ref{Eq:EOM}).  The orbital frequency $\omega_s$ is related to the angular momentum by $R\omega_s$, so we get
\begin{align}
\omega_s^2 &= \frac{GM_2}{R^2 r_1} = \dfrac{GM_{\text{tot}}}{R^3} ,
\label{Eq:orbital_frequency}
\end{align}
where $M_{\rm tot}=M_1+M_2$. Moreover, we incorporate the effects of GW back-reaction and dynamical friction into the dynamics of the SMBH orbit by considering the energy balance. For inspiraling SMBHBs, a part of its energy $E_{\text{orbit}}$ is converted into GW emission loss $E_{\text{gw}}$ and dynamical friction loss $E_{\text{df}}$. Thus the following energy balance equation is satisfied:
\begin{align}
 -\dfrac{{\rm d} E_{\text{orbit}}}{{\rm d}t} = \dfrac{{\rm d} E_{\text{gw}}}{{\rm d}t} + \dfrac{{\rm d} E_{\text{df}}}{{\rm d} t}. \label{Eq:energy_balance_eq}
\end{align}
As we will see in this subsection, this energy balance equation gives the time evolution of the orbital radius. The resulting orbit can be regarded as a quasi-circular orbit because of the smallness of these dissipative effects.

The orbital energy $E_{\text{orbit}}$ is the sum of the kinetic energy and the gravitational potential of the stellar mass object, so we can calculate $E_{\text{orbit}}$ using Eq. (\ref{Eq:orbital_frequency}),

\begin{equation}
\begin{aligned}
E_{\text{orbit}}
&= \frac{1}{2}M_1 \omega_s^2 r_1^2 + \frac{1}{2}M_2 \omega_s^2 r_2^2 -\frac{GM_1M_2}{R} \\
&= -\frac{GM_{\rm tot}\mu}{2R}, 
\label{Eq:orbital_energy} 
\end{aligned}
\end{equation}
where $v$ is the orbital velocity. When we consider the evolution of the radius $R$, $dR/dt$ does not vanish. So the time derivative of Eq. (\ref{Eq:orbital_energy}) gives the following equation,
\begin{equation}
\dfrac{{\rm d} E_{\text{orbit}}}{{\rm d}t} = \dfrac{GM_{\text{tot}} \mu}{2R^2} \dfrac{{\rm d}R}{{\rm d}t}. 
\label{Eq:E_orbit}
\end{equation}
To the lowest order in the Post Newtonian expansion, the gravitational radiation energy is given by the quadrupole formula.  We apply the formula to the circular Newtonian binary and obtain
\begin{equation}
 \dfrac{{\rm d} E_{\text{gw}}}{{\rm d}t} = \dfrac{32}{5} \dfrac{G\mu^2}{c^5} R^4 \omega_s^6 .  \label{Eq:E_GW}
\end{equation}

Based on the analysis detailed in Sec~\ref{sec:dynafric}, one can parameterize the expression for the dynamical friction acting on the SMBH binary as follows::
\begin{equation}
\begin{aligned}
\dfrac{{\rm d} E_{\text{df}}}{{\rm d}t} &= v f_{\text{df}} = 4\pi G^2\ln \Lambda \left[\frac{M_1^2}{\omega_s r_1}\rho_{\text{sp}}(r_1) + \frac{M_2^2}{\omega_s r_2}\rho_{\text{sp}}(r_2) \right] \\
&= 4\pi G^2\ln \Lambda \frac{\mu^2 R^2}{\omega_s} \left[\frac{\rho_{\text{sp}}(r_1)}{r_1^3} + \frac{\rho_{\text{sp}}(r_2)}{r_2^3} \right].
\label{Eq:E_DF}
\end{aligned}
\end{equation}
where the $\ln \Lambda$ is the parameter related to the galactic material, and we take $\ln \Lambda = 3$ in our analysis.

\subsection{SGWB Spectrum}\label{sec:SGWB_spectrum}
In the detector reference frame, the GW waveform from the binary composed of two compact objects is given by~\citep{Eda:2014kra}
\begin{subequations}
\begin{align}
 h_{+}\left(t\right)
  &= \dfrac{1}{d_{\rm M}} \dfrac{4G\mu \omega_s(t)^2 R(t)^2}{c^4} \dfrac{1 + \cos ^2 \iota}{2} \cos\left[ \Phi(t)\right], \label{Eq:GWwaveform_plus_including_backreaction} \\
 h_{\times}\left(t\right)
  &= \dfrac{1}{d_{\rm M}} \dfrac{4G\mu \omega_s(t)^2 R(t)^2}{c^4}  \cos \iota \sin\left[ \Phi(t)\right], \label{Eq:GWwaveform_cross_including_backreaction} \\
 \Phi \left(t\right)
  &= \int^t \omega_{\rm gw} \left(t'\right) \ {\rm d}t'. \label{Eq:phase_t_def}
\end{align}
\end{subequations}
where 
$d_{\rm M}$ is the proper-motion distance to the source,
$R$ is the distance between two BHs, $\iota$ is the inclination angle, which is the angle between the line-of-sight and the rotational axis of the orbits, and $\omega_{\rm gw}$ is the GW frequency which is given by $\omega_{\rm gw}(t) \equiv 2 \omega_s (\tau(t))/(1+z)$.
$A\left( t\right)$ is the time-dependent amplitude and
$\Phi\left(t\right)$ is the time-dependent GW phase.
In the frequency domain, the GW waveform is given by
\begin{align}
 \tilde{h}_{+,\times}  \left(f\right) = \int_{-\infty}^{\infty} h_{+, \times} \left(t\right) e^{-2\pi ift} {\rm d}t,
\end{align}
where $f$ is the GW frequency.  
In the range of frequency that we are concerned with, the time-dependent amplitude $A\left(t\right)$ varies slowly, while the time-dependent phase $\Phi\left(t\right)$ varies rapidly.  So, the Fourier transform of the GW waveform can be calculated approximately using the stationary phase method.  In this method, the rapidly oscillating term is neglected and only the slowly oscillating term survives. Then the GW waveform in the Fourier domain becomes
\begin{subequations}
\begin{align}
&\left|\tilde{h}_{+}(f)\right| = \dfrac{1}{2d_{\rm M}} \dfrac{4G\mu \omega_s^2(t) R^2(t)}{c^4} \sqrt{\frac{2\pi}{\ddot{\Phi}(t)}} \dfrac{1 + \cos ^2 \iota}{2}, \label{Eq:waveformf1}\\
&\left|\tilde{h}_{\times}(f)\right| = \dfrac{1}{2d_{\rm M}} \dfrac{4G\mu \omega_s^2(t) R^2(t)}{c^4}  \sqrt{\frac{2\pi}{\ddot{\Phi}(t)}} \cos \iota, \label{Eq:waveformf2}
\end{align}
\end{subequations}
where  the time $t$ is related to frequency by $2\pi f = \omega_{\text{gw}} \left(t\right)$.
In the Newtonian limit, a circular binary of component masses $M_1$ and $M_2$ which merges 
in less than the age of the universe has~\citep{Phinney:2001di}
\begin{equation}
\begin{aligned}
\frac{{\rm d} E_{\rm gw}}{{\rm d} f_{r}}&=\frac{2 \pi^{2} c^{3}}{G} d_{\rm M}^{2} f^{2}\left\langle\left|\tilde{h}_{+}(f)\right|^{2}+\left|\tilde{h}_{\times}(f)\right|^{2}\right\rangle_{\Omega_{s}}
 \\ &= \frac{\pi}{3G} \frac{(G\mathcal{M})^{5/3}}{(\pi f_r)^{1/3}} \frac{{\rm d}E_{\rm gw}/{\rm d}t}{{\rm d}E_{\rm gw}/{\rm d}t + {\rm d}E_{\rm df}/{\rm d}t},
\end{aligned}
\end{equation}
where $\mathcal{M}=M_{\rm tot}^{2/5} \mu^{3/5}$ is the chirp mass of the SMBH system.

The stochastic background is described in terms of the present-day GW energy density $\rho_{\rm gw}(f)c^2$ per logarithmic frequency interval normalized to the critical energy density $\rho_\mathrm{c}$ according to:
\begin{equation}
\Omega_\mathrm{gw}(f) \equiv \frac{1}{\rho_\mathrm{c}}\,\frac{ {\rm d} \rho_{\rm gw}(f)}{{\rm d} \ln f}={\pi \over 4}\,\frac{f^2h_c^2(f)}{G \rho_{\rm c}},
\label{e:Omega}
\end{equation}
where $f$ is the observed frequency in the detector reference frame.
It is contributed by the superposition of the radiation from the cosmic events, therefore the characteristic GW amplitude in a logarithmic frequency interval $h_{\rm c}$ is given by~\citep{Phinney:2001di}
\begin{equation}
h_c^2(f) =\frac{4G}{\pi f c^2}\iiint 
{\rm d}z {\rm d}M_{\rm tot} {\rm d}\mu \, \frac{{\rm d}^2n}{{\rm d}z{\rm d}{\mathcal{M}}}
~{{{\rm d}E_{\rm gw}(f; {M_{\rm tot}, \mu, z})} \over {{\rm d}{f_r}}},
\label{hcdE}
\end{equation}
where $f_r = (1 + z) f$ is the frequency measured in the source rest-frame at redshift $z$.
We use a parameterized population model similar to that of~\citep{Chen:2016kax,Chen:2016zyo,Chen:2018znx}
\begin{eqnarray}
    \frac{{\rm d}^2 n}{{\rm d}z {\rm d} \lg\mathcal{M}} =& \nonumber 
    \no \left[ \left( \frac{\mathcal{M}}{10^7\Msol}\right)^{-\alpha} e^{-\mathcal{M}/\mathcal{M}_*} e^{-z/z_0} \right] \\
    & \times (1+z)^{\betaz}  
    \frac{{\rm d}t_{\rm R}}{{\rm d}z},
    \label{eqn:model}
\end{eqnarray}
where $t_{\rm R}$ is the time in the rest frame of the source. 
The population model is described by five parameters $\{\no,\alpha,\mathcal{M}_*,\betaz,z_0\}$.
It is important to point out that there is another parameter $\gamma_{\rm sp}$ for the DM spike model that describes the DM density slope as detailed in Sec.~\ref{sec:spike}.

\section{Data Analysis and Results}\label{sec:data_analysis}

\begin{figure}[htbp!]
\centering
\includegraphics[width=0.917\linewidth]{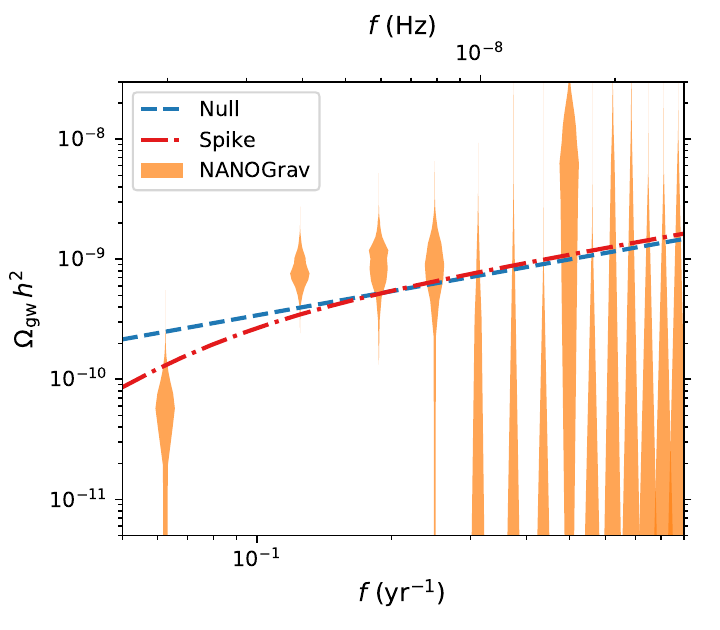}
\includegraphics[width=0.917\linewidth]{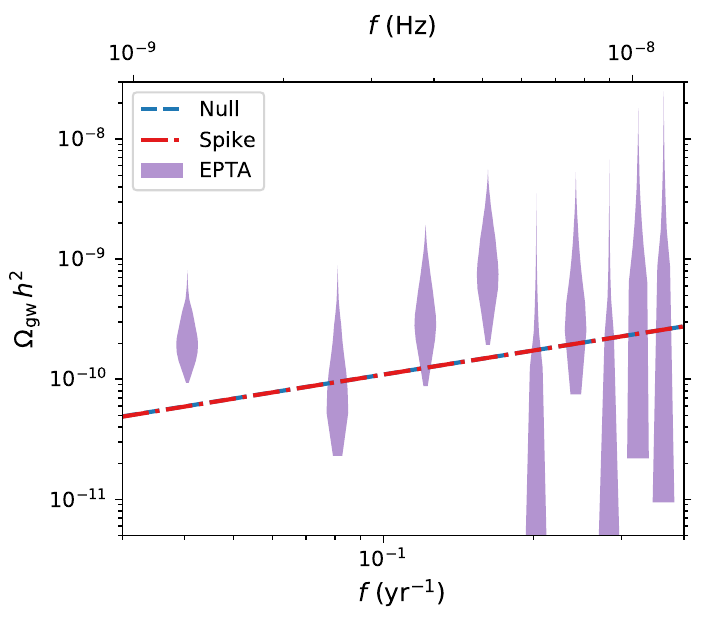}
\includegraphics[width=0.917\linewidth]{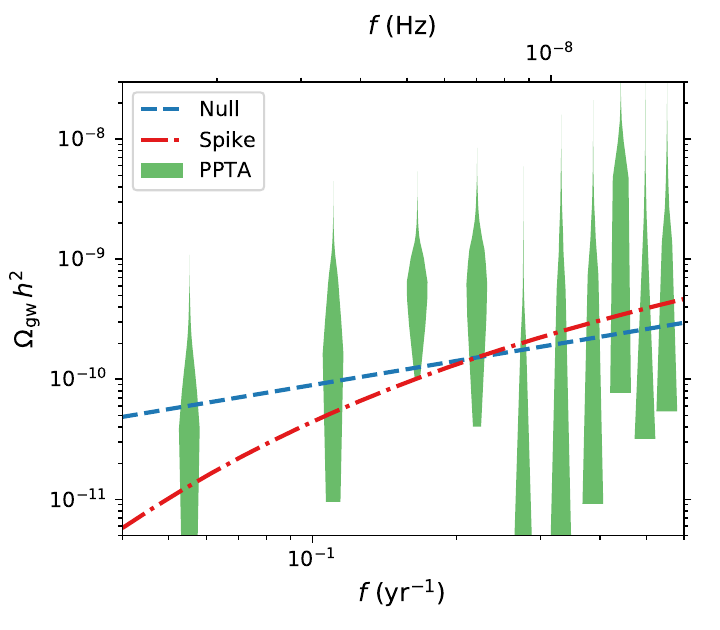}
\caption{The different SGWB spectra are displayed.
The upper, middle and lower panels show the observations from NANOGrav 15yr~\cite{NANOGrav:2023gor}, EPTA DR2~\cite{Antoniadis:2023ott}, and PPTA DR3~\cite{Reardon:2023gzh}, respectively.
The blue dashed line and red dot-dashed line are the best-fit values of cases with and without dark matter spike clothing.}
\label{fig_Omega_h2}
\end{figure}

The PTA collaborations have released their measurements on the Hellings-Downs-correlated common noise. Utilizing the most recent posteriors of the SGWB spectra from NANOGrav 15yr~\cite{NANOGrav:2023gor}, EPTA DR2~\cite{Antoniadis:2023ott}, and PPTA DR3~\cite{Reardon:2023gzh}, we employ Bayesian inference to identify the optimal fit for the SGWB spectra and investigate the constraints and implications on the DM spike around SMBHs.
For each frequency $f_i$, we calculate the posterior probability density function $\mathcal{P}_j(\Omega_{\rm gw}(f_i)h^2)$ of the GW energy density using the Monte Carlo chains from the PTA experiment (denoted with label $j$).
The likelihood function for the GW model $\Omega_{\mathrm{gw}}^{\rm model}(f_{i}; \Theta)$, either GW-only model or GW+DM spike model, is as follows:
\begin{equation}
\mathcal{L}_j(\Theta)=\prod_{i=1}^{N_{{\rm freq},j}} \mathcal{P}_{j}\left(\Omega_{\mathrm{gw}}^{\rm model}\left(f_{i}; \Theta\right)h^2\right).
\end{equation}
The free parameters in the energy density model are $\Theta = \{ \lg\left[\dot{n}_0/({\rm Mpc^{-3}\,Gyr^{-1}})\right], \lg (\mathcal{M}_{*}/M_\odot), z_0, \alpha, \beta_z, \gamma_{\rm sp} \}$.
We adopt uniform priors for the parameters, whose ranges are listed in the second column of the Table~\ref{tab::results}.
The Markov Chain Monte Carlo (MCMC) sampler \texttt{emcee}~\cite{emcee} are utilized to sample the posterior distribution and find the optimal values of the parameters.

\begin{table*}[!tbp]
  \centering
  \footnotesize
  \caption{\label{tab::results}
      The median and the 68\% credible regions of the parameters derived from each PTA experiment.
      The last row shows the maximum likelihood values of the models.
  }
  \begin{tabular}{c|c|c|c|c|c|c|c}
   \hline\hline
   \multirow{2}{*}{Parameter} & \multirow{2}{*}{Prior} & \multicolumn{2}{c|}{NANOGrav} & \multicolumn{2}{c|}{EPTA} & \multicolumn{2}{c}{PPTA}  \\
    \cline{3-8}
   \multirow{2}{*}{} & \multirow{2}{*}{} & Null & Spike & Null & Spike & Null & Spike  \\
    \hline
    $\lg (\dot{n}_0/({\rm Mpc^{-3}\,Gyr^{-1})})$
                         & $\mathcal{U}(-13,7)$   & $-3.0^{+1.8}_{-3.8}$ & $-4.3^{+2.7}_{-4.9}$ & $-3.3^{+1.7}_{-3.4}$ & $-5.0^{+2.6}_{-4.4}$ & $-3.4^{+1.8}_{-3.5}$ & $-3.6^{+2.1}_{-5.7}$ \\ 
    $\lg (\mathcal{M}_\star/M_\odot)$
                         & $\mathcal{U}(6,10)$    & $8.0^{+1.4}_{-1.4}$  & $8.5^{+1.0}_{-1.6}$  & $7.9^{+1.4}_{-1.3}$  & $8.4^{+1.0}_{-1.2}$  & $7.9^{+1.4}_{-1.3}$  & $7.9^{+1.3}_{-1.0}$ \\ 
    $z_0$                & $\mathcal{U}(0.1,4)$   & $2.1^{+1.3}_{-1.3}$  & $2.1^{+1.4}_{-1.6}$  & $2.0^{+1.3}_{-1.3}$  & $2.1^{+1.2}_{-1.3}$  & $2.0^{+1.3}_{-1.3}$  & $2.1^{+1.3}_{-1.2}$ \\ 
    $\alpha$             & $\mathcal{U}(-3,3)$    & $0.0^{+2.1}_{-1.9}$  & $-1.4^{+2.2}_{-1.2}$ & $0.2^{+1.9}_{-2.1}$  & $-0.8^{+1.7}_{-1.5}$ & $0.1^{+2.0}_{-2.0}$  & $-0.2^{+1.9}_{-1.8}$ \\ 
    $\beta_z$            & $\mathcal{U}(-2,7)$    & $2.5^{+3.1}_{-3.1}$  & $3.2^{+2.1}_{-2.9}$  & $2.5^{+3.0}_{-3.0}$  & $3.4^{+2.6}_{-3.7}$  & $2.5^{+3.0}_{-3.0}$  & $2.5^{+2.7}_{-3.0}$ \\ 
    $\gamma_{\rm sp}$    & $\mathcal{U}(0,3)$     &      $\cdots$        & $0.7^{+0.4}_{-0.4}$  &         $\cdots$     & $0.5^{+0.4}_{-0.3}$  &      $\cdots$        & $0.8^{+0.7}_{-0.5}$ \\
    \hline
    \multicolumn{2}{c|}{$-2\ln \mathcal{L}$}      & $-527.86$            & {$-532.11$}    & $-366.71$            & {$-366.72$}    & $-403.70$             & {$-405.01$} \\
    \hline\hline
    \end{tabular}
\end{table*}

In Table~\ref{tab::results}, we present the median and 68\% credible regions of the parameters.
For the SMBHB model without DM (null model), only the merger rate $\dot{n}_0$ is constrained by the data.
The optimal null model is shown with the blue dashed line in Fig.~\ref{fig_Omega_h2}.
Once we consider the DM surrounding the SMBHs, the fitting improves in the NANOGrav and PPTA data sets.
We illustrate the optimal models with the red dot-dashed line in Fig.~\ref{fig_Omega_h2}.
The dynamical friction decreases the GW strains at lower frequencies since the effect consumes more energy in outer orbits than gravitational wave.
Interestingly, the median value of the DM spike slope $\gamma_{\rm sp}$ is 0.7 in the NANOGrav data set with the 68\% credible region between 0.3 and 1.1, suggesting the spike may be flattened during the inspiral of the SMBHBs.

\section{Conclusions}\label{sec:conclusion}
DM spike is a steep structure enhancement formed during the adiabatic growth of the SMBHs~\citep{Gondolo:1999ef}, which can influence the dynamics of SMBH binaries through dynamical friction. In this work, we investigated the impact of such DM spikes on the spectrum of the SGWB, focusing on their potential to alter the GW spectrum at nanohertz frequencies. We show that the presence of a DM spike can lead to a flattening of the SGWB spectrum at low frequencies. As a result, measurements of the SGWB can be used to constrain the slope of the DM spike.

To test this hypothesis, we fitted the DM spike model to the recent PTA observations from NANOGrav~\cite{NANOGrav:2023gor}, EPTA~\cite{Antoniadis:2023ott}, and PPTA~\cite{Reardon:2023gzh}.
This model performs slightly better than that without DM for the NANOGrav and PPTA data sets as shown in the Table~\ref{tab::results}.
Interestingly, the best-fit value of the spike slope $\gamma_{\rm sp}$ is found to be smaller than 1, which may suggest that the spike is flattened around the SMBHBs.

With ongoing improvements in timing precision and increased sensitivity from PTA experiments worldwide, future observations of the SGWB may enable the reconstruction of the DM density profile around merging SMBHBs. This could open a new avenue for probing the nature of DM through gravitational-wave astronomy.

\vspace{0.5cm}

\begin{acknowledgements}
We appreciate Lei Lei, Yaoyu Li, Shao-Jiang Wang, Zi-Qing Xia, Qiang Yuan and Chi Zhang for helpful discussion. This work is supported by the National Key Research and Development Program of China (No. 2022YFF0503304), the National Natural Science Foundation of China (No. 12003074), and the New Corner stone Science Foundation through the XPLORER PRIZE. G.-W.Y. acknowledge support from the Istituto Nazionale di Fisica Nucleare (INFN) through the Commissione Scientifica Nazionale 4 (CSN4) Iniziativa Specifica “Quantum Fields in Gravity, Cosmology and Black Holes” (FLAG), and the Autonomous Province of Trento through the UniTrento Internal Call for Research 2023 grant “Searching for Dark Energy off the beaten track” (DARKTRACK, grant agreement No. E63C22000500003).
\end{acknowledgements}

\vspace{0.2cm}
\noindent \textit{Corresponding author}: G.-W. Yuan (guanwen.yuan@unitn.it);  Y.-Z. Fan (yzfan@pmo.ac.cn).


\begin{thebibliography}{127}%
    \makeatletter
    \providecommand \@ifxundefined [1]{%
     \@ifx{#1\undefined}
    }%
    \providecommand \@ifnum [1]{%
     \ifnum #1\expandafter \@firstoftwo
     \else \expandafter \@secondoftwo
     \fi
    }%
    \providecommand \@ifx [1]{%
     \ifx #1\expandafter \@firstoftwo
     \else \expandafter \@secondoftwo
     \fi
    }%
    \providecommand \natexlab [1]{#1}%
    \providecommand \enquote  [1]{``#1''}%
    \providecommand \bibnamefont  [1]{#1}%
    \providecommand \bibfnamefont [1]{#1}%
    \providecommand \citenamefont [1]{#1}%
    \providecommand \href@noop [0]{\@secondoftwo}%
    \providecommand \href [0]{\begingroup \@sanitize@url \@href}%
    \providecommand \@href[1]{\@@startlink{#1}\@@href}%
    \providecommand \@@href[1]{\endgroup#1\@@endlink}%
    \providecommand \@sanitize@url [0]{\catcode `\\12\catcode `\$12\catcode
      `\&12\catcode `\#12\catcode `\^12\catcode `\_12\catcode `\%12\relax}%
    \providecommand \@@startlink[1]{}%
    \providecommand \@@endlink[0]{}%
    \providecommand \url  [0]{\begingroup\@sanitize@url \@url }%
    \providecommand \@url [1]{\endgroup\@href {#1}{\urlprefix }}%
    \providecommand \urlprefix  [0]{URL }%
    \providecommand \Eprint [0]{\href }%
    \providecommand \doibase [0]{http://dx.doi.org/}%
    \providecommand \selectlanguage [0]{\@gobble}%
    \providecommand \bibinfo  [0]{\@secondoftwo}%
    \providecommand \bibfield  [0]{\@secondoftwo}%
    \providecommand \translation [1]{[#1]}%
    \providecommand \BibitemOpen [0]{}%
    \providecommand \bibitemStop [0]{}%
    \providecommand \bibitemNoStop [0]{.\EOS\space}%
    \providecommand \EOS [0]{\spacefactor3000\relax}%
    \providecommand \BibitemShut  [1]{\csname bibitem#1\endcsname}%
    \let\auto@bib@innerbib\@empty
    \bibitem [{\citenamefont {Weisberg}\ and\ \citenamefont
      {Taylor}(2005)}]{Weisberg:2004hi}%
      \BibitemOpen
      \bibfield  {author} {\bibinfo {author} {\bibfnamefont {J.~M.}\ \bibnamefont
      {Weisberg}}\ and\ \bibinfo {author} {\bibfnamefont {J.~H.}\ \bibnamefont
      {Taylor}},\ }\href@noop {} {\bibfield  {journal} {\bibinfo  {journal} {ASP
      Conf. Ser.}\ }\textbf {\bibinfo {volume} {328}},\ \bibinfo {pages} {25}
      (\bibinfo {year} {2005})},\ \Eprint {http://arxiv.org/abs/astro-ph/0407149}
      {arXiv:astro-ph/0407149} \BibitemShut {NoStop}%
    \bibitem [{\citenamefont {Abbott}\ \emph {et~al.}(2016)\citenamefont {Abbott}
      \emph {et~al.}}]{LIGOScientific:2016aoc}%
      \BibitemOpen
      \bibfield  {author} {\bibinfo {author} {\bibfnamefont {B.~P.}\ \bibnamefont
      {Abbott}} \emph {et~al.} (\bibinfo {collaboration} {LIGO Scientific,
      Virgo}),\ }\href {\doibase 10.1103/PhysRevLett.116.061102} {\bibfield
      {journal} {\bibinfo  {journal} {Phys. Rev. Lett.}\ }\textbf {\bibinfo
      {volume} {116}},\ \bibinfo {pages} {061102} (\bibinfo {year} {2016})},\
      \Eprint {http://arxiv.org/abs/1602.03837} {arXiv:1602.03837 [gr-qc]}
      \BibitemShut {NoStop}%
    \bibitem [{\citenamefont {Arzoumanian}\ \emph {et~al.}(2020)\citenamefont
      {Arzoumanian} \emph {et~al.}}]{NANOGrav:2020bcs}%
      \BibitemOpen
      \bibfield  {author} {\bibinfo {author} {\bibfnamefont {Z.}~\bibnamefont
      {Arzoumanian}} \emph {et~al.} (\bibinfo {collaboration} {NANOGrav}),\ }\href
      {\doibase 10.3847/2041-8213/abd401} {\bibfield  {journal} {\bibinfo
      {journal} {Astrophys. J. Lett.}\ }\textbf {\bibinfo {volume} {905}},\
      \bibinfo {pages} {L34} (\bibinfo {year} {2020})},\ \Eprint
      {http://arxiv.org/abs/2009.04496} {arXiv:2009.04496 [astro-ph.HE]}
      \BibitemShut {NoStop}%
    \bibitem [{\citenamefont {Chen}\ \emph {et~al.}(2021)\citenamefont {Chen} \emph
      {et~al.}}]{Chen:2021rqp}%
      \BibitemOpen
      \bibfield  {author} {\bibinfo {author} {\bibfnamefont {S.}~\bibnamefont
      {Chen}} \emph {et~al.} (\bibinfo {collaboration} {EPTA}),\ }\href {\doibase
      10.1093/mnras/stab2833} {\bibfield  {journal} {\bibinfo  {journal} {Mon. Not.
      Roy. Astron. Soc.}\ }\textbf {\bibinfo {volume} {508}},\ \bibinfo {pages}
      {4970} (\bibinfo {year} {2021})},\ \Eprint {http://arxiv.org/abs/2110.13184}
      {arXiv:2110.13184 [astro-ph.HE]} \BibitemShut {NoStop}%
    \bibitem [{\citenamefont {Goncharov}\ \emph {et~al.}(2021)\citenamefont
      {Goncharov} \emph {et~al.}}]{Goncharov:2021oub}%
      \BibitemOpen
      \bibfield  {author} {\bibinfo {author} {\bibfnamefont {B.}~\bibnamefont
      {Goncharov}} \emph {et~al.},\ }\href {\doibase 10.3847/2041-8213/ac17f4}
      {\bibfield  {journal} {\bibinfo  {journal} {Astrophys. J. Lett.}\ }\textbf
      {\bibinfo {volume} {917}},\ \bibinfo {pages} {L19} (\bibinfo {year}
      {2021})},\ \Eprint {http://arxiv.org/abs/2107.12112} {arXiv:2107.12112
      [astro-ph.HE]} \BibitemShut {NoStop}%
    \bibitem [{\citenamefont {Agazie}\ \emph
      {et~al.}(2023{\natexlab{a}})\citenamefont {Agazie} \emph
      {et~al.}}]{NANOGrav:2023gor}%
      \BibitemOpen
      \bibfield  {author} {\bibinfo {author} {\bibfnamefont {G.}~\bibnamefont
      {Agazie}} \emph {et~al.} (\bibinfo {collaboration} {NANOGrav}),\ }\href
      {\doibase 10.3847/2041-8213/acdac6} {\bibfield  {journal} {\bibinfo
      {journal} {Astrophys. J. Lett.}\ }\textbf {\bibinfo {volume} {951}},\
      \bibinfo {pages} {L8} (\bibinfo {year} {2023}{\natexlab{a}})},\ \Eprint
      {http://arxiv.org/abs/2306.16213} {arXiv:2306.16213 [astro-ph.HE]}
      \BibitemShut {NoStop}%
    \bibitem [{\citenamefont {Antoniadis}\ \emph {et~al.}(2023)\citenamefont
      {Antoniadis} \emph {et~al.}}]{Antoniadis:2023ott}%
      \BibitemOpen
      \bibfield  {author} {\bibinfo {author} {\bibfnamefont {J.}~\bibnamefont
      {Antoniadis}} \emph {et~al.} (\bibinfo {collaboration} {EPTA, InPTA}),\
      }\href {\doibase 10.1051/0004-6361/202346844} {\bibfield  {journal} {\bibinfo
       {journal} {Astron. Astrophys.}\ }\textbf {\bibinfo {volume} {678}},\
      \bibinfo {pages} {A50} (\bibinfo {year} {2023})},\ \Eprint
      {http://arxiv.org/abs/2306.16214} {arXiv:2306.16214 [astro-ph.HE]}
      \BibitemShut {NoStop}%
    \bibitem [{\citenamefont {Reardon}\ \emph {et~al.}(2023)\citenamefont {Reardon}
      \emph {et~al.}}]{Reardon:2023gzh}%
      \BibitemOpen
      \bibfield  {author} {\bibinfo {author} {\bibfnamefont {D.~J.}\ \bibnamefont
      {Reardon}} \emph {et~al.},\ }\href {\doibase 10.3847/2041-8213/acdd02}
      {\bibfield  {journal} {\bibinfo  {journal} {Astrophys. J. Lett.}\ }\textbf
      {\bibinfo {volume} {951}},\ \bibinfo {pages} {L6} (\bibinfo {year} {2023})},\
      \Eprint {http://arxiv.org/abs/2306.16215} {arXiv:2306.16215 [astro-ph.HE]}
      \BibitemShut {NoStop}%
    \bibitem [{\citenamefont {Xu}\ \emph {et~al.}(2023)\citenamefont {Xu} \emph
      {et~al.}}]{Xu:2023wog}%
      \BibitemOpen
      \bibfield  {author} {\bibinfo {author} {\bibfnamefont {H.}~\bibnamefont {Xu}}
      \emph {et~al.},\ }\href {\doibase 10.1088/1674-4527/acdfa5} {\bibfield
      {journal} {\bibinfo  {journal} {Res. Astron. Astrophys.}\ }\textbf {\bibinfo
      {volume} {23}},\ \bibinfo {pages} {075024} (\bibinfo {year} {2023})},\
      \Eprint {http://arxiv.org/abs/2306.16216} {arXiv:2306.16216 [astro-ph.HE]}
      \BibitemShut {NoStop}%
    \bibitem [{\citenamefont {{Miles}}\ \emph {et~al.}(2025)\citenamefont
      {{Miles}}, \citenamefont {{Shannon}}, \citenamefont {{Reardon}},
      \citenamefont {{Bailes}} \emph {et~al.}}]{2025MNRAS.536.1467M}%
      \BibitemOpen
      \bibfield  {author} {\bibinfo {author} {\bibfnamefont {M.~T.}\ \bibnamefont
      {{Miles}}}, \bibinfo {author} {\bibfnamefont {R.~M.}\ \bibnamefont
      {{Shannon}}}, \bibinfo {author} {\bibfnamefont {D.~J.}\ \bibnamefont
      {{Reardon}}}, \bibinfo {author} {\bibfnamefont {M.}~\bibnamefont {{Bailes}}},
       \emph {et~al.},\ }\href {\doibase 10.1093/mnras/stae2572} {\bibfield
      {journal} {\bibinfo  {journal} {\mnras}\ }\textbf {\bibinfo {volume} {536}},\
      \bibinfo {pages} {1467} (\bibinfo {year} {2025})},\ \Eprint
      {http://arxiv.org/abs/2412.01148} {arXiv:2412.01148 [astro-ph.HE]}
      \BibitemShut {NoStop}%
    \bibitem [{\citenamefont {Afzal}\ \emph {et~al.}(2023)\citenamefont {Afzal}
      \emph {et~al.}}]{NANOGrav:2023hvm}%
      \BibitemOpen
      \bibfield  {author} {\bibinfo {author} {\bibfnamefont {A.}~\bibnamefont
      {Afzal}} \emph {et~al.} (\bibinfo {collaboration} {NANOGrav}),\ }\href
      {\doibase 10.3847/2041-8213/acdc91} {\bibfield  {journal} {\bibinfo
      {journal} {Astrophys. J. Lett.}\ }\textbf {\bibinfo {volume} {951}},\
      \bibinfo {pages} {L11} (\bibinfo {year} {2023})},\ \bibinfo {note} {[Erratum:
      Astrophys.J.Lett. 971, L27 (2024), Erratum: Astrophys.J. 971, L27 (2024)]},\
      \Eprint {http://arxiv.org/abs/2306.16219} {arXiv:2306.16219 [astro-ph.HE]}
      \BibitemShut {NoStop}%
    \bibitem [{\citenamefont {Antoniadis}\ \emph {et~al.}(2024)\citenamefont
      {Antoniadis} \emph {et~al.}}]{Antoniadis:2023xlr}%
      \BibitemOpen
      \bibfield  {author} {\bibinfo {author} {\bibfnamefont {J.}~\bibnamefont
      {Antoniadis}} \emph {et~al.} (\bibinfo {collaboration} {EPTA, InPTA}),\
      }\href {\doibase 10.1051/0004-6361/202347433} {\bibfield  {journal} {\bibinfo
       {journal} {Astron. Astrophys.}\ }\textbf {\bibinfo {volume} {685}},\
      \bibinfo {pages} {A94} (\bibinfo {year} {2024})},\ \Eprint
      {http://arxiv.org/abs/2306.16227} {arXiv:2306.16227 [astro-ph.CO]}
      \BibitemShut {NoStop}%
    \bibitem [{\citenamefont {Burke-Spolaor}\ \emph {et~al.}(2019)\citenamefont
      {Burke-Spolaor} \emph {et~al.}}]{Burke-Spolaor:2018bvk}%
      \BibitemOpen
      \bibfield  {author} {\bibinfo {author} {\bibfnamefont {S.}~\bibnamefont
      {Burke-Spolaor}} \emph {et~al.},\ }\href {\doibase 10.1007/s00159-019-0115-7}
      {\bibfield  {journal} {\bibinfo  {journal} {Astron. Astrophys. Rev.}\
      }\textbf {\bibinfo {volume} {27}},\ \bibinfo {pages} {5} (\bibinfo {year}
      {2019})},\ \Eprint {http://arxiv.org/abs/1811.08826} {arXiv:1811.08826
      [astro-ph.HE]} \BibitemShut {NoStop}%
    \bibitem [{\citenamefont {Sesana}\ \emph
      {et~al.}(2008{\natexlab{a}})\citenamefont {Sesana}, \citenamefont {Vecchio},\
      and\ \citenamefont {Colacino}}]{Sesana:2008mz}%
      \BibitemOpen
      \bibfield  {author} {\bibinfo {author} {\bibfnamefont {A.}~\bibnamefont
      {Sesana}}, \bibinfo {author} {\bibfnamefont {A.}~\bibnamefont {Vecchio}}, \
      and\ \bibinfo {author} {\bibfnamefont {C.~N.}\ \bibnamefont {Colacino}},\
      }\href {\doibase 10.1111/j.1365-2966.2008.13682.x} {\bibfield  {journal}
      {\bibinfo  {journal} {Mon. Not. Roy. Astron. Soc.}\ }\textbf {\bibinfo
      {volume} {390}},\ \bibinfo {pages} {192} (\bibinfo {year}
      {2008}{\natexlab{a}})},\ \Eprint {http://arxiv.org/abs/0804.4476}
      {arXiv:0804.4476 [astro-ph]} \BibitemShut {NoStop}%
    \bibitem [{\citenamefont {Broadhurst}\ \emph {et~al.}(2023)\citenamefont
      {Broadhurst}, \citenamefont {Chen}, \citenamefont {Liu},\ and\ \citenamefont
      {Zheng}}]{Broadhurst:2023tus}%
      \BibitemOpen
      \bibfield  {author} {\bibinfo {author} {\bibfnamefont {T.}~\bibnamefont
      {Broadhurst}}, \bibinfo {author} {\bibfnamefont {C.}~\bibnamefont {Chen}},
      \bibinfo {author} {\bibfnamefont {T.}~\bibnamefont {Liu}}, \ and\ \bibinfo
      {author} {\bibfnamefont {K.-F.}\ \bibnamefont {Zheng}},\ }\href@noop {} {\
      (\bibinfo {year} {2023})},\ \Eprint {http://arxiv.org/abs/2306.17821}
      {arXiv:2306.17821 [astro-ph.HE]} \BibitemShut {NoStop}%
    \bibitem [{\citenamefont {Agazie}\ \emph
      {et~al.}(2023{\natexlab{b}})\citenamefont {Agazie} \emph
      {et~al.}}]{NANOGrav:2023hfp}%
      \BibitemOpen
      \bibfield  {author} {\bibinfo {author} {\bibfnamefont {G.}~\bibnamefont
      {Agazie}} \emph {et~al.} (\bibinfo {collaboration} {NANOGrav}),\ }\href
      {\doibase 10.3847/2041-8213/ace18b} {\bibfield  {journal} {\bibinfo
      {journal} {Astrophys. J. Lett.}\ }\textbf {\bibinfo {volume} {952}},\
      \bibinfo {pages} {L37} (\bibinfo {year} {2023}{\natexlab{b}})},\ \Eprint
      {http://arxiv.org/abs/2306.16220} {arXiv:2306.16220 [astro-ph.HE]}
      \BibitemShut {NoStop}%
    \bibitem [{\citenamefont {Agazie}\ \emph
      {et~al.}(2023{\natexlab{c}})\citenamefont {Agazie} \emph
      {et~al.}}]{NANOGrav:2023pdq}%
      \BibitemOpen
      \bibfield  {author} {\bibinfo {author} {\bibfnamefont {G.}~\bibnamefont
      {Agazie}} \emph {et~al.} (\bibinfo {collaboration} {NANOGrav}),\ }\href
      {\doibase 10.3847/2041-8213/ace18a} {\bibfield  {journal} {\bibinfo
      {journal} {Astrophys. J. Lett.}\ }\textbf {\bibinfo {volume} {951}},\
      \bibinfo {pages} {L50} (\bibinfo {year} {2023}{\natexlab{c}})},\ \Eprint
      {http://arxiv.org/abs/2306.16222} {arXiv:2306.16222 [astro-ph.HE]}
      \BibitemShut {NoStop}%
    \bibitem [{\citenamefont {Franciolini}\ \emph {et~al.}(2024)\citenamefont
      {Franciolini}, \citenamefont {Racco},\ and\ \citenamefont
      {Rompineve}}]{Franciolini:2023wjm}%
      \BibitemOpen
      \bibfield  {author} {\bibinfo {author} {\bibfnamefont {G.}~\bibnamefont
      {Franciolini}}, \bibinfo {author} {\bibfnamefont {D.}~\bibnamefont {Racco}},
      \ and\ \bibinfo {author} {\bibfnamefont {F.}~\bibnamefont {Rompineve}},\
      }\href {\doibase 10.1103/PhysRevLett.132.081001} {\bibfield  {journal}
      {\bibinfo  {journal} {Phys. Rev. Lett.}\ }\textbf {\bibinfo {volume} {132}},\
      \bibinfo {pages} {081001} (\bibinfo {year} {2024})},\ \bibinfo {note}
      {[Erratum: Phys.Rev.Lett. 133, 189901 (2024)]},\ \Eprint
      {http://arxiv.org/abs/2306.17136} {arXiv:2306.17136 [astro-ph.CO]}
      \BibitemShut {NoStop}%
    \bibitem [{\citenamefont {Lambiase}\ \emph {et~al.}(2023)\citenamefont
      {Lambiase}, \citenamefont {Mastrototaro},\ and\ \citenamefont
      {Visinelli}}]{Lambiase:2023pxd}%
      \BibitemOpen
      \bibfield  {author} {\bibinfo {author} {\bibfnamefont {G.}~\bibnamefont
      {Lambiase}}, \bibinfo {author} {\bibfnamefont {L.}~\bibnamefont
      {Mastrototaro}}, \ and\ \bibinfo {author} {\bibfnamefont {L.}~\bibnamefont
      {Visinelli}},\ }\href {\doibase 10.1103/PhysRevD.108.123028} {\bibfield
      {journal} {\bibinfo  {journal} {Phys. Rev. D}\ }\textbf {\bibinfo {volume}
      {108}},\ \bibinfo {pages} {123028} (\bibinfo {year} {2023})},\ \Eprint
      {http://arxiv.org/abs/2306.16977} {arXiv:2306.16977 [astro-ph.HE]}
      \BibitemShut {NoStop}%
    \bibitem [{\citenamefont {Han}\ \emph {et~al.}(2024)\citenamefont {Han},
      \citenamefont {Xie}, \citenamefont {Yang},\ and\ \citenamefont
      {Zhang}}]{Han:2023olf}%
      \BibitemOpen
      \bibfield  {author} {\bibinfo {author} {\bibfnamefont {C.}~\bibnamefont
      {Han}}, \bibinfo {author} {\bibfnamefont {K.-P.}\ \bibnamefont {Xie}},
      \bibinfo {author} {\bibfnamefont {J.~M.}\ \bibnamefont {Yang}}, \ and\
      \bibinfo {author} {\bibfnamefont {M.}~\bibnamefont {Zhang}},\ }\href
      {\doibase 10.1103/PhysRevD.109.115025} {\bibfield  {journal} {\bibinfo
      {journal} {Phys. Rev. D}\ }\textbf {\bibinfo {volume} {109}},\ \bibinfo
      {pages} {115025} (\bibinfo {year} {2024})},\ \Eprint
      {http://arxiv.org/abs/2306.16966} {arXiv:2306.16966 [hep-ph]} \BibitemShut
      {NoStop}%
    \bibitem [{\citenamefont {Guo}\ \emph {et~al.}(2024)\citenamefont {Guo},
      \citenamefont {Khlopov}, \citenamefont {Liu}, \citenamefont {Wu},
      \citenamefont {Wu},\ and\ \citenamefont {Zhu}}]{Guo:2023hyp}%
      \BibitemOpen
      \bibfield  {author} {\bibinfo {author} {\bibfnamefont {S.-Y.}\ \bibnamefont
      {Guo}}, \bibinfo {author} {\bibfnamefont {M.}~\bibnamefont {Khlopov}},
      \bibinfo {author} {\bibfnamefont {X.}~\bibnamefont {Liu}}, \bibinfo {author}
      {\bibfnamefont {L.}~\bibnamefont {Wu}}, \bibinfo {author} {\bibfnamefont
      {Y.}~\bibnamefont {Wu}}, \ and\ \bibinfo {author} {\bibfnamefont
      {B.}~\bibnamefont {Zhu}},\ }\href {\doibase 10.1007/s11433-024-2445-1}
      {\bibfield  {journal} {\bibinfo  {journal} {Sci. China Phys. Mech. Astron.}\
      }\textbf {\bibinfo {volume} {67}},\ \bibinfo {pages} {111011} (\bibinfo
      {year} {2024})},\ \Eprint {http://arxiv.org/abs/2306.17022} {arXiv:2306.17022
      [hep-ph]} \BibitemShut {NoStop}%
    \bibitem [{\citenamefont {Wang}\ \emph {et~al.}(2023)\citenamefont {Wang},
      \citenamefont {Lei}, \citenamefont {Jiao}, \citenamefont {Feng},\ and\
      \citenamefont {Fan}}]{Wang:2023len}%
      \BibitemOpen
      \bibfield  {author} {\bibinfo {author} {\bibfnamefont {Z.}~\bibnamefont
      {Wang}}, \bibinfo {author} {\bibfnamefont {L.}~\bibnamefont {Lei}}, \bibinfo
      {author} {\bibfnamefont {H.}~\bibnamefont {Jiao}}, \bibinfo {author}
      {\bibfnamefont {L.}~\bibnamefont {Feng}}, \ and\ \bibinfo {author}
      {\bibfnamefont {Y.-Z.}\ \bibnamefont {Fan}},\ }\href {\doibase
      10.1007/s11433-023-2262-0} {\bibfield  {journal} {\bibinfo  {journal} {Sci.
      China Phys. Mech. Astron.}\ }\textbf {\bibinfo {volume} {66}},\ \bibinfo
      {pages} {120403} (\bibinfo {year} {2023})},\ \Eprint
      {http://arxiv.org/abs/2306.17150} {arXiv:2306.17150 [astro-ph.HE]}
      \BibitemShut {NoStop}%
    \bibitem [{\citenamefont {Ellis}\ \emph {et~al.}(2023)\citenamefont {Ellis},
      \citenamefont {Lewicki}, \citenamefont {Lin},\ and\ \citenamefont
      {Vaskonen}}]{Ellis:2023tsl}%
      \BibitemOpen
      \bibfield  {author} {\bibinfo {author} {\bibfnamefont {J.}~\bibnamefont
      {Ellis}}, \bibinfo {author} {\bibfnamefont {M.}~\bibnamefont {Lewicki}},
      \bibinfo {author} {\bibfnamefont {C.}~\bibnamefont {Lin}}, \ and\ \bibinfo
      {author} {\bibfnamefont {V.}~\bibnamefont {Vaskonen}},\ }\href {\doibase
      10.1103/PhysRevD.108.103511} {\bibfield  {journal} {\bibinfo  {journal}
      {Phys. Rev. D}\ }\textbf {\bibinfo {volume} {108}},\ \bibinfo {pages}
      {103511} (\bibinfo {year} {2023})},\ \Eprint
      {http://arxiv.org/abs/2306.17147} {arXiv:2306.17147 [astro-ph.CO]}
      \BibitemShut {NoStop}%
    \bibitem [{\citenamefont {Vagnozzi}(2023)}]{Vagnozzi:2023lwo}%
      \BibitemOpen
      \bibfield  {author} {\bibinfo {author} {\bibfnamefont {S.}~\bibnamefont
      {Vagnozzi}},\ }\href {\doibase 10.1016/j.jheap.2023.07.001} {\bibfield
      {journal} {\bibinfo  {journal} {JHEAp}\ }\textbf {\bibinfo {volume} {39}},\
      \bibinfo {pages} {81} (\bibinfo {year} {2023})},\ \Eprint
      {http://arxiv.org/abs/2306.16912} {arXiv:2306.16912 [astro-ph.CO]}
      \BibitemShut {NoStop}%
    \bibitem [{\citenamefont {Liu}\ \emph {et~al.}(2024)\citenamefont {Liu},
      \citenamefont {Chen},\ and\ \citenamefont {Huang}}]{Liu:2023ymk}%
      \BibitemOpen
      \bibfield  {author} {\bibinfo {author} {\bibfnamefont {L.}~\bibnamefont
      {Liu}}, \bibinfo {author} {\bibfnamefont {Z.-C.}\ \bibnamefont {Chen}}, \
      and\ \bibinfo {author} {\bibfnamefont {Q.-G.}\ \bibnamefont {Huang}},\ }\href
      {\doibase 10.1103/PhysRevD.109.L061301} {\bibfield  {journal} {\bibinfo
      {journal} {Phys. Rev. D}\ }\textbf {\bibinfo {volume} {109}},\ \bibinfo
      {pages} {L061301} (\bibinfo {year} {2024})},\ \Eprint
      {http://arxiv.org/abs/2307.01102} {arXiv:2307.01102 [astro-ph.CO]}
      \BibitemShut {NoStop}%
    \bibitem [{\citenamefont {Wu}\ \emph {et~al.}(2024)\citenamefont {Wu},
      \citenamefont {Chen},\ and\ \citenamefont {Huang}}]{Wu:2023hsa}%
      \BibitemOpen
      \bibfield  {author} {\bibinfo {author} {\bibfnamefont {Y.-M.}\ \bibnamefont
      {Wu}}, \bibinfo {author} {\bibfnamefont {Z.-C.}\ \bibnamefont {Chen}}, \ and\
      \bibinfo {author} {\bibfnamefont {Q.-G.}\ \bibnamefont {Huang}},\ }\href
      {\doibase 10.1007/s11433-023-2298-7} {\bibfield  {journal} {\bibinfo
      {journal} {Sci. China Phys. Mech. Astron.}\ }\textbf {\bibinfo {volume}
      {67}},\ \bibinfo {pages} {240412} (\bibinfo {year} {2024})},\ \Eprint
      {http://arxiv.org/abs/2307.03141} {arXiv:2307.03141 [astro-ph.CO]}
      \BibitemShut {NoStop}%
    \bibitem [{\citenamefont {Fujikura}\ \emph {et~al.}(2023)\citenamefont
      {Fujikura}, \citenamefont {Girmohanta}, \citenamefont {Nakai},\ and\
      \citenamefont {Suzuki}}]{Fujikura:2023lkn}%
      \BibitemOpen
      \bibfield  {author} {\bibinfo {author} {\bibfnamefont {K.}~\bibnamefont
      {Fujikura}}, \bibinfo {author} {\bibfnamefont {S.}~\bibnamefont
      {Girmohanta}}, \bibinfo {author} {\bibfnamefont {Y.}~\bibnamefont {Nakai}}, \
      and\ \bibinfo {author} {\bibfnamefont {M.}~\bibnamefont {Suzuki}},\ }\href
      {\doibase 10.1016/j.physletb.2023.138203} {\bibfield  {journal} {\bibinfo
      {journal} {Phys. Lett. B}\ }\textbf {\bibinfo {volume} {846}},\ \bibinfo
      {pages} {138203} (\bibinfo {year} {2023})},\ \Eprint
      {http://arxiv.org/abs/2306.17086} {arXiv:2306.17086 [hep-ph]} \BibitemShut
      {NoStop}%
    \bibitem [{\citenamefont {Kitajima}\ \emph {et~al.}(2024)\citenamefont
      {Kitajima}, \citenamefont {Lee}, \citenamefont {Murai}, \citenamefont
      {Takahashi},\ and\ \citenamefont {Yin}}]{Kitajima:2023cek}%
      \BibitemOpen
      \bibfield  {author} {\bibinfo {author} {\bibfnamefont {N.}~\bibnamefont
      {Kitajima}}, \bibinfo {author} {\bibfnamefont {J.}~\bibnamefont {Lee}},
      \bibinfo {author} {\bibfnamefont {K.}~\bibnamefont {Murai}}, \bibinfo
      {author} {\bibfnamefont {F.}~\bibnamefont {Takahashi}}, \ and\ \bibinfo
      {author} {\bibfnamefont {W.}~\bibnamefont {Yin}},\ }\href {\doibase
      10.1016/j.physletb.2024.138586} {\bibfield  {journal} {\bibinfo  {journal}
      {Phys. Lett. B}\ }\textbf {\bibinfo {volume} {851}},\ \bibinfo {pages}
      {138586} (\bibinfo {year} {2024})},\ \Eprint
      {http://arxiv.org/abs/2306.17146} {arXiv:2306.17146 [hep-ph]} \BibitemShut
      {NoStop}%
    \bibitem [{\citenamefont {Franciolini}\ \emph {et~al.}(2023)\citenamefont
      {Franciolini}, \citenamefont {Iovino}, \citenamefont {Vaskonen},\ and\
      \citenamefont {Veermae}}]{Franciolini:2023pbf}%
      \BibitemOpen
      \bibfield  {author} {\bibinfo {author} {\bibfnamefont {G.}~\bibnamefont
      {Franciolini}}, \bibinfo {author} {\bibfnamefont {A.}~\bibnamefont {Iovino},
      \bibfnamefont {Junior.}}, \bibinfo {author} {\bibfnamefont {V.}~\bibnamefont
      {Vaskonen}}, \ and\ \bibinfo {author} {\bibfnamefont {H.}~\bibnamefont
      {Veermae}},\ }\href {\doibase 10.1103/PhysRevLett.131.201401} {\bibfield
      {journal} {\bibinfo  {journal} {Phys. Rev. Lett.}\ }\textbf {\bibinfo
      {volume} {131}},\ \bibinfo {pages} {201401} (\bibinfo {year} {2023})},\
      \Eprint {http://arxiv.org/abs/2306.17149} {arXiv:2306.17149 [astro-ph.CO]}
      \BibitemShut {NoStop}%
    \bibitem [{\citenamefont {Megias}\ \emph {et~al.}(2023)\citenamefont {Megias},
      \citenamefont {Nardini},\ and\ \citenamefont {Quiros}}]{Megias:2023kiy}%
      \BibitemOpen
      \bibfield  {author} {\bibinfo {author} {\bibfnamefont {E.}~\bibnamefont
      {Megias}}, \bibinfo {author} {\bibfnamefont {G.}~\bibnamefont {Nardini}}, \
      and\ \bibinfo {author} {\bibfnamefont {M.}~\bibnamefont {Quiros}},\ }\href
      {\doibase 10.1103/PhysRevD.108.095017} {\bibfield  {journal} {\bibinfo
      {journal} {Phys. Rev. D}\ }\textbf {\bibinfo {volume} {108}},\ \bibinfo
      {pages} {095017} (\bibinfo {year} {2023})},\ \Eprint
      {http://arxiv.org/abs/2306.17071} {arXiv:2306.17071 [hep-ph]} \BibitemShut
      {NoStop}%
    \bibitem [{\citenamefont {Ellis}\ \emph
      {et~al.}(2024{\natexlab{a}})\citenamefont {Ellis}, \citenamefont {Fairbairn},
      \citenamefont {H\"utsi}, \citenamefont {Raidal}, \citenamefont {Urrutia},
      \citenamefont {Vaskonen},\ and\ \citenamefont {Veerm\"ae}}]{Ellis:2023dgf}%
      \BibitemOpen
      \bibfield  {author} {\bibinfo {author} {\bibfnamefont {J.}~\bibnamefont
      {Ellis}}, \bibinfo {author} {\bibfnamefont {M.}~\bibnamefont {Fairbairn}},
      \bibinfo {author} {\bibfnamefont {G.}~\bibnamefont {H\"utsi}}, \bibinfo
      {author} {\bibfnamefont {J.}~\bibnamefont {Raidal}}, \bibinfo {author}
      {\bibfnamefont {J.}~\bibnamefont {Urrutia}}, \bibinfo {author} {\bibfnamefont
      {V.}~\bibnamefont {Vaskonen}}, \ and\ \bibinfo {author} {\bibfnamefont
      {H.}~\bibnamefont {Veerm\"ae}},\ }\href {\doibase
      10.1103/PhysRevD.109.L021302} {\bibfield  {journal} {\bibinfo  {journal}
      {Phys. Rev. D}\ }\textbf {\bibinfo {volume} {109}},\ \bibinfo {pages}
      {L021302} (\bibinfo {year} {2024}{\natexlab{a}})},\ \Eprint
      {http://arxiv.org/abs/2306.17021} {arXiv:2306.17021 [astro-ph.CO]}
      \BibitemShut {NoStop}%
    \bibitem [{\citenamefont {Bai}\ \emph {et~al.}(2023)\citenamefont {Bai},
      \citenamefont {Chen},\ and\ \citenamefont {Korwar}}]{Bai:2023cqj}%
      \BibitemOpen
      \bibfield  {author} {\bibinfo {author} {\bibfnamefont {Y.}~\bibnamefont
      {Bai}}, \bibinfo {author} {\bibfnamefont {T.-K.}\ \bibnamefont {Chen}}, \
      and\ \bibinfo {author} {\bibfnamefont {M.}~\bibnamefont {Korwar}},\ }\href
      {\doibase 10.1007/JHEP12(2023)194} {\bibfield  {journal} {\bibinfo  {journal}
      {JHEP}\ }\textbf {\bibinfo {volume} {12}},\ \bibinfo {pages} {194} (\bibinfo
      {year} {2023})},\ \Eprint {http://arxiv.org/abs/2306.17160} {arXiv:2306.17160
      [hep-ph]} \BibitemShut {NoStop}%
    \bibitem [{\citenamefont {Yang}\ \emph {et~al.}(2024)\citenamefont {Yang},
      \citenamefont {Xie},\ and\ \citenamefont {Huang}}]{Yang:2023aak}%
      \BibitemOpen
      \bibfield  {author} {\bibinfo {author} {\bibfnamefont {J.}~\bibnamefont
      {Yang}}, \bibinfo {author} {\bibfnamefont {N.}~\bibnamefont {Xie}}, \ and\
      \bibinfo {author} {\bibfnamefont {F.~P.}\ \bibnamefont {Huang}},\ }\href
      {\doibase 10.1088/1475-7516/2024/11/045} {\bibfield  {journal} {\bibinfo
      {journal} {JCAP}\ }\textbf {\bibinfo {volume} {11}},\ \bibinfo {pages} {045}
      (\bibinfo {year} {2024})},\ \Eprint {http://arxiv.org/abs/2306.17113}
      {arXiv:2306.17113 [hep-ph]} \BibitemShut {NoStop}%
    \bibitem [{\citenamefont {Ghoshal}\ and\ \citenamefont
      {Strumia}(2024)}]{Ghoshal:2023fhh}%
      \BibitemOpen
      \bibfield  {author} {\bibinfo {author} {\bibfnamefont {A.}~\bibnamefont
      {Ghoshal}}\ and\ \bibinfo {author} {\bibfnamefont {A.}~\bibnamefont
      {Strumia}},\ }\href {\doibase 10.1088/1475-7516/2024/02/054} {\bibfield
      {journal} {\bibinfo  {journal} {JCAP}\ }\textbf {\bibinfo {volume} {02}},\
      \bibinfo {pages} {054} (\bibinfo {year} {2024})},\ \Eprint
      {http://arxiv.org/abs/2306.17158} {arXiv:2306.17158 [astro-ph.CO]}
      \BibitemShut {NoStop}%
    \bibitem [{\citenamefont {Deng}\ \emph {et~al.}(2023)\citenamefont {Deng},
      \citenamefont {B\'ecsy}, \citenamefont {Siemens}, \citenamefont {Cornish},\
      and\ \citenamefont {Madison}}]{Deng:2023btv}%
      \BibitemOpen
      \bibfield  {author} {\bibinfo {author} {\bibfnamefont {H.}~\bibnamefont
      {Deng}}, \bibinfo {author} {\bibfnamefont {B.}~\bibnamefont {B\'ecsy}},
      \bibinfo {author} {\bibfnamefont {X.}~\bibnamefont {Siemens}}, \bibinfo
      {author} {\bibfnamefont {N.~J.}\ \bibnamefont {Cornish}}, \ and\ \bibinfo
      {author} {\bibfnamefont {D.~R.}\ \bibnamefont {Madison}},\ }\href {\doibase
      10.1103/PhysRevD.108.102007} {\bibfield  {journal} {\bibinfo  {journal}
      {Phys. Rev. D}\ }\textbf {\bibinfo {volume} {108}},\ \bibinfo {pages}
      {102007} (\bibinfo {year} {2023})},\ \Eprint
      {http://arxiv.org/abs/2306.17130} {arXiv:2306.17130 [gr-qc]} \BibitemShut
      {NoStop}%
    \bibitem [{\citenamefont {Mitridate}\ \emph {et~al.}(2023)\citenamefont
      {Mitridate}, \citenamefont {Wright}, \citenamefont {von Eckardstein},
      \citenamefont {Schr\"oder}, \citenamefont {Nay}, \citenamefont {Olum},
      \citenamefont {Schmitz},\ and\ \citenamefont {Trickle}}]{Mitridate:2023oar}%
      \BibitemOpen
      \bibfield  {author} {\bibinfo {author} {\bibfnamefont {A.}~\bibnamefont
      {Mitridate}}, \bibinfo {author} {\bibfnamefont {D.}~\bibnamefont {Wright}},
      \bibinfo {author} {\bibfnamefont {R.}~\bibnamefont {von Eckardstein}},
      \bibinfo {author} {\bibfnamefont {T.}~\bibnamefont {Schr\"oder}}, \bibinfo
      {author} {\bibfnamefont {J.}~\bibnamefont {Nay}}, \bibinfo {author}
      {\bibfnamefont {K.}~\bibnamefont {Olum}}, \bibinfo {author} {\bibfnamefont
      {K.}~\bibnamefont {Schmitz}}, \ and\ \bibinfo {author} {\bibfnamefont
      {T.}~\bibnamefont {Trickle}},\ }\href@noop {} {\  (\bibinfo {year} {2023})},\
      \Eprint {http://arxiv.org/abs/2306.16377} {arXiv:2306.16377 [hep-ph]}
      \BibitemShut {NoStop}%
    \bibitem [{\citenamefont {King}\ \emph {et~al.}(2024)\citenamefont {King},
      \citenamefont {Marfatia},\ and\ \citenamefont {Rahat}}]{King:2023cgv}%
      \BibitemOpen
      \bibfield  {author} {\bibinfo {author} {\bibfnamefont {S.~F.}\ \bibnamefont
      {King}}, \bibinfo {author} {\bibfnamefont {D.}~\bibnamefont {Marfatia}}, \
      and\ \bibinfo {author} {\bibfnamefont {M.~H.}\ \bibnamefont {Rahat}},\ }\href
      {\doibase 10.1103/PhysRevD.109.035014} {\bibfield  {journal} {\bibinfo
      {journal} {Phys. Rev. D}\ }\textbf {\bibinfo {volume} {109}},\ \bibinfo
      {pages} {035014} (\bibinfo {year} {2024})},\ \Eprint
      {http://arxiv.org/abs/2306.05389} {arXiv:2306.05389 [hep-ph]} \BibitemShut
      {NoStop}%
    \bibitem [{\citenamefont {Zu}\ \emph {et~al.}(2024)\citenamefont {Zu},
      \citenamefont {Zhang}, \citenamefont {Li}, \citenamefont {Gu}, \citenamefont
      {Tsai},\ and\ \citenamefont {Fan}}]{Zu:2023olm}%
      \BibitemOpen
      \bibfield  {author} {\bibinfo {author} {\bibfnamefont {L.}~\bibnamefont
      {Zu}}, \bibinfo {author} {\bibfnamefont {C.}~\bibnamefont {Zhang}}, \bibinfo
      {author} {\bibfnamefont {Y.-Y.}\ \bibnamefont {Li}}, \bibinfo {author}
      {\bibfnamefont {Y.}~\bibnamefont {Gu}}, \bibinfo {author} {\bibfnamefont
      {Y.-L.~S.}\ \bibnamefont {Tsai}}, \ and\ \bibinfo {author} {\bibfnamefont
      {Y.-Z.}\ \bibnamefont {Fan}},\ }\href {\doibase 10.1016/j.scib.2024.01.037}
      {\bibfield  {journal} {\bibinfo  {journal} {Sci. Bull.}\ }\textbf {\bibinfo
      {volume} {69}},\ \bibinfo {pages} {741} (\bibinfo {year} {2024})},\ \Eprint
      {http://arxiv.org/abs/2306.16769} {arXiv:2306.16769 [astro-ph.HE]}
      \BibitemShut {NoStop}%
    \bibitem [{\citenamefont {Li}\ \emph {et~al.}(2024)\citenamefont {Li},
      \citenamefont {Zhang}, \citenamefont {Wang}, \citenamefont {Cui},
      \citenamefont {Tsai}, \citenamefont {Yuan},\ and\ \citenamefont
      {Fan}}]{Li:2023yaj}%
      \BibitemOpen
      \bibfield  {author} {\bibinfo {author} {\bibfnamefont {Y.-Y.}\ \bibnamefont
      {Li}}, \bibinfo {author} {\bibfnamefont {C.}~\bibnamefont {Zhang}}, \bibinfo
      {author} {\bibfnamefont {Z.}~\bibnamefont {Wang}}, \bibinfo {author}
      {\bibfnamefont {M.-Y.}\ \bibnamefont {Cui}}, \bibinfo {author} {\bibfnamefont
      {Y.-L.~S.}\ \bibnamefont {Tsai}}, \bibinfo {author} {\bibfnamefont
      {Q.}~\bibnamefont {Yuan}}, \ and\ \bibinfo {author} {\bibfnamefont {Y.-Z.}\
      \bibnamefont {Fan}},\ }\href {\doibase 10.1103/PhysRevD.109.043538}
      {\bibfield  {journal} {\bibinfo  {journal} {Phys. Rev. D}\ }\textbf {\bibinfo
      {volume} {109}},\ \bibinfo {pages} {043538} (\bibinfo {year} {2024})},\
      \Eprint {http://arxiv.org/abs/2306.17124} {arXiv:2306.17124 [astro-ph.HE]}
      \BibitemShut {NoStop}%
    \bibitem [{\citenamefont {Addazi}\ \emph {et~al.}(2024)\citenamefont {Addazi},
      \citenamefont {Cai}, \citenamefont {Marciano},\ and\ \citenamefont
      {Visinelli}}]{Addazi:2023jvg}%
      \BibitemOpen
      \bibfield  {author} {\bibinfo {author} {\bibfnamefont {A.}~\bibnamefont
      {Addazi}}, \bibinfo {author} {\bibfnamefont {Y.-F.}\ \bibnamefont {Cai}},
      \bibinfo {author} {\bibfnamefont {A.}~\bibnamefont {Marciano}}, \ and\
      \bibinfo {author} {\bibfnamefont {L.}~\bibnamefont {Visinelli}},\ }\href
      {\doibase 10.1103/PhysRevD.109.015028} {\bibfield  {journal} {\bibinfo
      {journal} {Phys. Rev. D}\ }\textbf {\bibinfo {volume} {109}},\ \bibinfo
      {pages} {015028} (\bibinfo {year} {2024})},\ \Eprint
      {http://arxiv.org/abs/2306.17205} {arXiv:2306.17205 [astro-ph.CO]}
      \BibitemShut {NoStop}%
    \bibitem [{\citenamefont {Konoplya}\ and\ \citenamefont
      {Zhidenko}(2024)}]{Konoplya:2023fmh}%
      \BibitemOpen
      \bibfield  {author} {\bibinfo {author} {\bibfnamefont {R.~A.}\ \bibnamefont
      {Konoplya}}\ and\ \bibinfo {author} {\bibfnamefont {A.}~\bibnamefont
      {Zhidenko}},\ }\href {\doibase 10.1016/j.physletb.2024.138685} {\bibfield
      {journal} {\bibinfo  {journal} {Phys. Lett. B}\ }\textbf {\bibinfo {volume}
      {853}},\ \bibinfo {pages} {138685} (\bibinfo {year} {2024})},\ \Eprint
      {http://arxiv.org/abs/2307.01110} {arXiv:2307.01110 [gr-qc]} \BibitemShut
      {NoStop}%
    \bibitem [{\citenamefont {Unal}\ \emph {et~al.}(2024)\citenamefont {Unal},
      \citenamefont {Papageorgiou},\ and\ \citenamefont {Obata}}]{Unal:2023srk}%
      \BibitemOpen
      \bibfield  {author} {\bibinfo {author} {\bibfnamefont {C.}~\bibnamefont
      {Unal}}, \bibinfo {author} {\bibfnamefont {A.}~\bibnamefont {Papageorgiou}},
      \ and\ \bibinfo {author} {\bibfnamefont {I.}~\bibnamefont {Obata}},\ }\href
      {\doibase 10.1016/j.physletb.2024.138873} {\bibfield  {journal} {\bibinfo
      {journal} {Phys. Lett. B}\ }\textbf {\bibinfo {volume} {856}},\ \bibinfo
      {pages} {138873} (\bibinfo {year} {2024})},\ \Eprint
      {http://arxiv.org/abs/2307.02322} {arXiv:2307.02322 [astro-ph.CO]}
      \BibitemShut {NoStop}%
    \bibitem [{\citenamefont {Bian}\ \emph
      {et~al.}(2024{\natexlab{a}})\citenamefont {Bian}, \citenamefont {Ge},
      \citenamefont {Shu}, \citenamefont {Wang}, \citenamefont {Yang},\ and\
      \citenamefont {Zong}}]{Bian:2023dnv}%
      \BibitemOpen
      \bibfield  {author} {\bibinfo {author} {\bibfnamefont {L.}~\bibnamefont
      {Bian}}, \bibinfo {author} {\bibfnamefont {S.}~\bibnamefont {Ge}}, \bibinfo
      {author} {\bibfnamefont {J.}~\bibnamefont {Shu}}, \bibinfo {author}
      {\bibfnamefont {B.}~\bibnamefont {Wang}}, \bibinfo {author} {\bibfnamefont
      {X.-Y.}\ \bibnamefont {Yang}}, \ and\ \bibinfo {author} {\bibfnamefont
      {J.}~\bibnamefont {Zong}},\ }\href {\doibase 10.1103/PhysRevD.109.L101301}
      {\bibfield  {journal} {\bibinfo  {journal} {Phys. Rev. D}\ }\textbf {\bibinfo
      {volume} {109}},\ \bibinfo {pages} {L101301} (\bibinfo {year}
      {2024}{\natexlab{a}})},\ \Eprint {http://arxiv.org/abs/2307.02376}
      {arXiv:2307.02376 [astro-ph.HE]} \BibitemShut {NoStop}%
    \bibitem [{\citenamefont {Madge}\ \emph
      {et~al.}(2023{\natexlab{a}})\citenamefont {Madge}, \citenamefont {Morgante},
      \citenamefont {Puchades-Ib\'a\~nez}, \citenamefont {Ramberg}, \citenamefont
      {Ratzinger}, \citenamefont {Schenk},\ and\ \citenamefont
      {Schwaller}}]{Madge:2023cak}%
      \BibitemOpen
      \bibfield  {author} {\bibinfo {author} {\bibfnamefont {E.}~\bibnamefont
      {Madge}}, \bibinfo {author} {\bibfnamefont {E.}~\bibnamefont {Morgante}},
      \bibinfo {author} {\bibfnamefont {C.}~\bibnamefont {Puchades-Ib\'a\~nez}},
      \bibinfo {author} {\bibfnamefont {N.}~\bibnamefont {Ramberg}}, \bibinfo
      {author} {\bibfnamefont {W.}~\bibnamefont {Ratzinger}}, \bibinfo {author}
      {\bibfnamefont {S.}~\bibnamefont {Schenk}}, \ and\ \bibinfo {author}
      {\bibfnamefont {P.}~\bibnamefont {Schwaller}},\ }\href {\doibase
      10.1007/JHEP10(2023)171} {\bibfield  {journal} {\bibinfo  {journal} {JHEP}\
      }\textbf {\bibinfo {volume} {10}},\ \bibinfo {pages} {171} (\bibinfo {year}
      {2023}{\natexlab{a}})},\ \Eprint {http://arxiv.org/abs/2306.14856}
      {arXiv:2306.14856 [hep-ph]} \BibitemShut {NoStop}%
    \bibitem [{\citenamefont {Balaji}\ \emph {et~al.}(2023)\citenamefont {Balaji},
      \citenamefont {Dom\`enech},\ and\ \citenamefont
      {Franciolini}}]{Balaji:2023ehk}%
      \BibitemOpen
      \bibfield  {author} {\bibinfo {author} {\bibfnamefont {S.}~\bibnamefont
      {Balaji}}, \bibinfo {author} {\bibfnamefont {G.}~\bibnamefont {Dom\`enech}},
      \ and\ \bibinfo {author} {\bibfnamefont {G.}~\bibnamefont {Franciolini}},\
      }\href {\doibase 10.1088/1475-7516/2023/10/041} {\bibfield  {journal}
      {\bibinfo  {journal} {JCAP}\ }\textbf {\bibinfo {volume} {10}},\ \bibinfo
      {pages} {041} (\bibinfo {year} {2023})},\ \Eprint
      {http://arxiv.org/abs/2307.08552} {arXiv:2307.08552 [gr-qc]} \BibitemShut
      {NoStop}%
    \bibitem [{\citenamefont {Liu}\ \emph {et~al.}(2023)\citenamefont {Liu},
      \citenamefont {Chen},\ and\ \citenamefont {Huang}}]{Liu:2023pau}%
      \BibitemOpen
      \bibfield  {author} {\bibinfo {author} {\bibfnamefont {L.}~\bibnamefont
      {Liu}}, \bibinfo {author} {\bibfnamefont {Z.-C.}\ \bibnamefont {Chen}}, \
      and\ \bibinfo {author} {\bibfnamefont {Q.-G.}\ \bibnamefont {Huang}},\ }\href
      {\doibase 10.1088/1475-7516/2023/11/071} {\bibfield  {journal} {\bibinfo
      {journal} {JCAP}\ }\textbf {\bibinfo {volume} {11}},\ \bibinfo {pages} {071}
      (\bibinfo {year} {2023})},\ \Eprint {http://arxiv.org/abs/2307.14911}
      {arXiv:2307.14911 [astro-ph.CO]} \BibitemShut {NoStop}%
    \bibitem [{\citenamefont {Madge}\ \emph
      {et~al.}(2023{\natexlab{b}})\citenamefont {Madge}, \citenamefont {Morgante},
      \citenamefont {Puchades-Ib\'a\~nez}, \citenamefont {Ramberg}, \citenamefont
      {Ratzinger}, \citenamefont {Schenk},\ and\ \citenamefont
      {Schwaller}}]{Madge:2023dxc}%
      \BibitemOpen
      \bibfield  {author} {\bibinfo {author} {\bibfnamefont {E.}~\bibnamefont
      {Madge}}, \bibinfo {author} {\bibfnamefont {E.}~\bibnamefont {Morgante}},
      \bibinfo {author} {\bibfnamefont {C.}~\bibnamefont {Puchades-Ib\'a\~nez}},
      \bibinfo {author} {\bibfnamefont {N.}~\bibnamefont {Ramberg}}, \bibinfo
      {author} {\bibfnamefont {W.}~\bibnamefont {Ratzinger}}, \bibinfo {author}
      {\bibfnamefont {S.}~\bibnamefont {Schenk}}, \ and\ \bibinfo {author}
      {\bibfnamefont {P.}~\bibnamefont {Schwaller}},\ }\href {\doibase
      10.1007/JHEP10(2023)171} {\bibfield  {journal} {\bibinfo  {journal} {JHEP}\
      }\textbf {\bibinfo {volume} {10}},\ \bibinfo {pages} {171} (\bibinfo {year}
      {2023}{\natexlab{b}})},\ \Eprint {http://arxiv.org/abs/2306.14856}
      {arXiv:2306.14856 [hep-ph]} \BibitemShut {NoStop}%
    \bibitem [{\citenamefont {Ellis}\ \emph
      {et~al.}(2024{\natexlab{b}})\citenamefont {Ellis}, \citenamefont {Fairbairn},
      \citenamefont {Franciolini}, \citenamefont {H\"utsi}, \citenamefont {Iovino},
      \citenamefont {Lewicki}, \citenamefont {Raidal}, \citenamefont {Urrutia},
      \citenamefont {Vaskonen},\ and\ \citenamefont {Veerm\"ae}}]{Ellis:2023oxs}%
      \BibitemOpen
      \bibfield  {author} {\bibinfo {author} {\bibfnamefont {J.}~\bibnamefont
      {Ellis}}, \bibinfo {author} {\bibfnamefont {M.}~\bibnamefont {Fairbairn}},
      \bibinfo {author} {\bibfnamefont {G.}~\bibnamefont {Franciolini}}, \bibinfo
      {author} {\bibfnamefont {G.}~\bibnamefont {H\"utsi}}, \bibinfo {author}
      {\bibfnamefont {A.}~\bibnamefont {Iovino}}, \bibinfo {author} {\bibfnamefont
      {M.}~\bibnamefont {Lewicki}}, \bibinfo {author} {\bibfnamefont
      {M.}~\bibnamefont {Raidal}}, \bibinfo {author} {\bibfnamefont
      {J.}~\bibnamefont {Urrutia}}, \bibinfo {author} {\bibfnamefont
      {V.}~\bibnamefont {Vaskonen}}, \ and\ \bibinfo {author} {\bibfnamefont
      {H.}~\bibnamefont {Veerm\"ae}},\ }\href {\doibase
      10.1103/PhysRevD.109.023522} {\bibfield  {journal} {\bibinfo  {journal}
      {Phys. Rev. D}\ }\textbf {\bibinfo {volume} {109}},\ \bibinfo {pages}
      {023522} (\bibinfo {year} {2024}{\natexlab{b}})},\ \Eprint
      {http://arxiv.org/abs/2308.08546} {arXiv:2308.08546 [astro-ph.CO]}
      \BibitemShut {NoStop}%
    \bibitem [{\citenamefont {Liang}\ \emph {et~al.}(2024)\citenamefont {Liang},
      \citenamefont {Lin}, \citenamefont {Trodden},\ and\ \citenamefont
      {Wong}}]{Liang:2023pbj}%
      \BibitemOpen
      \bibfield  {author} {\bibinfo {author} {\bibfnamefont {Q.}~\bibnamefont
      {Liang}}, \bibinfo {author} {\bibfnamefont {M.-X.}\ \bibnamefont {Lin}},
      \bibinfo {author} {\bibfnamefont {M.}~\bibnamefont {Trodden}}, \ and\
      \bibinfo {author} {\bibfnamefont {S.~S.~C.}\ \bibnamefont {Wong}},\ }\href
      {\doibase 10.1103/PhysRevD.109.083028} {\bibfield  {journal} {\bibinfo
      {journal} {Phys. Rev. D}\ }\textbf {\bibinfo {volume} {109}},\ \bibinfo
      {pages} {083028} (\bibinfo {year} {2024})},\ \Eprint
      {http://arxiv.org/abs/2309.16666} {arXiv:2309.16666 [astro-ph.CO]}
      \BibitemShut {NoStop}%
    \bibitem [{\citenamefont {Calz\`a}\ \emph {et~al.}(2024)\citenamefont
      {Calz\`a}, \citenamefont {Gianesello}, \citenamefont {Rinaldi},\ and\
      \citenamefont {Vagnozzi}}]{Calza:2024qxn}%
      \BibitemOpen
      \bibfield  {author} {\bibinfo {author} {\bibfnamefont {M.}~\bibnamefont
      {Calz\`a}}, \bibinfo {author} {\bibfnamefont {F.}~\bibnamefont {Gianesello}},
      \bibinfo {author} {\bibfnamefont {M.}~\bibnamefont {Rinaldi}}, \ and\
      \bibinfo {author} {\bibfnamefont {S.}~\bibnamefont {Vagnozzi}},\ }\href
      {\doibase 10.1038/s41598-024-82661-8} {\bibfield  {journal} {\bibinfo
      {journal} {Sci. Rep.}\ }\textbf {\bibinfo {volume} {14}},\ \bibinfo {pages}
      {31296} (\bibinfo {year} {2024})},\ \Eprint {http://arxiv.org/abs/2409.01801}
      {arXiv:2409.01801 [gr-qc]} \BibitemShut {NoStop}%
    \bibitem [{\citenamefont {Ivanov}\ \emph {et~al.}(1994)\citenamefont {Ivanov},
      \citenamefont {Naselsky},\ and\ \citenamefont {Novikov}}]{Ivanov:1994pa}%
      \BibitemOpen
      \bibfield  {author} {\bibinfo {author} {\bibfnamefont {P.}~\bibnamefont
      {Ivanov}}, \bibinfo {author} {\bibfnamefont {P.}~\bibnamefont {Naselsky}}, \
      and\ \bibinfo {author} {\bibfnamefont {I.}~\bibnamefont {Novikov}},\ }\href
      {\doibase 10.1103/PhysRevD.50.7173} {\bibfield  {journal} {\bibinfo
      {journal} {Phys. Rev. D}\ }\textbf {\bibinfo {volume} {50}},\ \bibinfo
      {pages} {7173} (\bibinfo {year} {1994})}\BibitemShut {NoStop}%
    \bibitem [{\citenamefont {Carr}\ \emph {et~al.}(2016)\citenamefont {Carr},
      \citenamefont {Kuhnel},\ and\ \citenamefont {Sandstad}}]{Carr:2016drx}%
      \BibitemOpen
      \bibfield  {author} {\bibinfo {author} {\bibfnamefont {B.}~\bibnamefont
      {Carr}}, \bibinfo {author} {\bibfnamefont {F.}~\bibnamefont {Kuhnel}}, \ and\
      \bibinfo {author} {\bibfnamefont {M.}~\bibnamefont {Sandstad}},\ }\href
      {\doibase 10.1103/PhysRevD.94.083504} {\bibfield  {journal} {\bibinfo
      {journal} {Phys. Rev. D}\ }\textbf {\bibinfo {volume} {94}},\ \bibinfo
      {pages} {083504} (\bibinfo {year} {2016})},\ \Eprint
      {http://arxiv.org/abs/1607.06077} {arXiv:1607.06077 [astro-ph.CO]}
      \BibitemShut {NoStop}%
    \bibitem [{\citenamefont {Carr}\ and\ \citenamefont
      {Kuhnel}(2020)}]{Carr:2020xqk}%
      \BibitemOpen
      \bibfield  {author} {\bibinfo {author} {\bibfnamefont {B.}~\bibnamefont
      {Carr}}\ and\ \bibinfo {author} {\bibfnamefont {F.}~\bibnamefont {Kuhnel}},\
      }\href {\doibase 10.1146/annurev-nucl-050520-125911} {\bibfield  {journal}
      {\bibinfo  {journal} {Ann. Rev. Nucl. Part. Sci.}\ }\textbf {\bibinfo
      {volume} {70}},\ \bibinfo {pages} {355} (\bibinfo {year} {2020})},\ \Eprint
      {http://arxiv.org/abs/2006.02838} {arXiv:2006.02838 [astro-ph.CO]}
      \BibitemShut {NoStop}%
    \bibitem [{\citenamefont {Chen}\ \emph {et~al.}(2020)\citenamefont {Chen},
      \citenamefont {Yuan},\ and\ \citenamefont {Huang}}]{Chen:2019xse}%
      \BibitemOpen
      \bibfield  {author} {\bibinfo {author} {\bibfnamefont {Z.-C.}\ \bibnamefont
      {Chen}}, \bibinfo {author} {\bibfnamefont {C.}~\bibnamefont {Yuan}}, \ and\
      \bibinfo {author} {\bibfnamefont {Q.-G.}\ \bibnamefont {Huang}},\ }\href
      {\doibase 10.1103/PhysRevLett.124.251101} {\bibfield  {journal} {\bibinfo
      {journal} {Phys. Rev. Lett.}\ }\textbf {\bibinfo {volume} {124}},\ \bibinfo
      {pages} {25} (\bibinfo {year} {2020})},\ \Eprint
      {http://arxiv.org/abs/1910.12239} {arXiv:1910.12239 [astro-ph.CO]}
      \BibitemShut {NoStop}%
    \bibitem [{\citenamefont {Yuan}\ \emph {et~al.}(2024)\citenamefont {Yuan},
      \citenamefont {Lei}, \citenamefont {Wang}, \citenamefont {Wang},
      \citenamefont {Wang}, \citenamefont {Chen}, \citenamefont {Shen},
      \citenamefont {Cai},\ and\ \citenamefont {Fan}}]{Yuan:2023bvh}%
      \BibitemOpen
      \bibfield  {author} {\bibinfo {author} {\bibfnamefont {G.-W.}\ \bibnamefont
      {Yuan}}, \bibinfo {author} {\bibfnamefont {L.}~\bibnamefont {Lei}}, \bibinfo
      {author} {\bibfnamefont {Y.-Z.}\ \bibnamefont {Wang}}, \bibinfo {author}
      {\bibfnamefont {B.}~\bibnamefont {Wang}}, \bibinfo {author} {\bibfnamefont
      {Y.-Y.}\ \bibnamefont {Wang}}, \bibinfo {author} {\bibfnamefont
      {C.}~\bibnamefont {Chen}}, \bibinfo {author} {\bibfnamefont {Z.-Q.}\
      \bibnamefont {Shen}}, \bibinfo {author} {\bibfnamefont {Y.-F.}\ \bibnamefont
      {Cai}}, \ and\ \bibinfo {author} {\bibfnamefont {Y.-Z.}\ \bibnamefont
      {Fan}},\ }\href {\doibase 10.1007/s11433-024-2433-3} {\bibfield  {journal}
      {\bibinfo  {journal} {Sci. China Phys. Mech. Astron.}\ }\textbf {\bibinfo
      {volume} {67}},\ \bibinfo {pages} {109512} (\bibinfo {year} {2024})},\
      \Eprint {http://arxiv.org/abs/2303.09391} {arXiv:2303.09391 [astro-ph.CO]}
      \BibitemShut {NoStop}%
    \bibitem [{\citenamefont {Wang}\ \emph {et~al.}(2024)\citenamefont {Wang},
      \citenamefont {Zhao}, \citenamefont {Li},\ and\ \citenamefont
      {Zhu}}]{Wang:2023ost}%
      \BibitemOpen
      \bibfield  {author} {\bibinfo {author} {\bibfnamefont {S.}~\bibnamefont
      {Wang}}, \bibinfo {author} {\bibfnamefont {Z.-C.}\ \bibnamefont {Zhao}},
      \bibinfo {author} {\bibfnamefont {J.-P.}\ \bibnamefont {Li}}, \ and\ \bibinfo
      {author} {\bibfnamefont {Q.-H.}\ \bibnamefont {Zhu}},\ }\href {\doibase
      10.1103/PhysRevResearch.6.L012060} {\bibfield  {journal} {\bibinfo  {journal}
      {Phys. Rev. Res.}\ }\textbf {\bibinfo {volume} {6}},\ \bibinfo {pages}
      {L012060} (\bibinfo {year} {2024})},\ \Eprint
      {http://arxiv.org/abs/2307.00572} {arXiv:2307.00572 [astro-ph.CO]}
      \BibitemShut {NoStop}%
    \bibitem [{\citenamefont {Inomata}\ \emph {et~al.}(2024)\citenamefont
      {Inomata}, \citenamefont {Kohri},\ and\ \citenamefont
      {Terada}}]{Inomata:2023zup}%
      \BibitemOpen
      \bibfield  {author} {\bibinfo {author} {\bibfnamefont {K.}~\bibnamefont
      {Inomata}}, \bibinfo {author} {\bibfnamefont {K.}~\bibnamefont {Kohri}}, \
      and\ \bibinfo {author} {\bibfnamefont {T.}~\bibnamefont {Terada}},\ }\href
      {\doibase 10.1103/PhysRevD.109.063506} {\bibfield  {journal} {\bibinfo
      {journal} {Phys. Rev. D}\ }\textbf {\bibinfo {volume} {109}},\ \bibinfo
      {pages} {063506} (\bibinfo {year} {2024})},\ \Eprint
      {http://arxiv.org/abs/2306.17834} {arXiv:2306.17834 [astro-ph.CO]}
      \BibitemShut {NoStop}%
    \bibitem [{\citenamefont {Figueroa}\ \emph {et~al.}(2024)\citenamefont
      {Figueroa}, \citenamefont {Pieroni}, \citenamefont {Ricciardone},\ and\
      \citenamefont {Simakachorn}}]{Figueroa:2023zhu}%
      \BibitemOpen
      \bibfield  {author} {\bibinfo {author} {\bibfnamefont {D.~G.}\ \bibnamefont
      {Figueroa}}, \bibinfo {author} {\bibfnamefont {M.}~\bibnamefont {Pieroni}},
      \bibinfo {author} {\bibfnamefont {A.}~\bibnamefont {Ricciardone}}, \ and\
      \bibinfo {author} {\bibfnamefont {P.}~\bibnamefont {Simakachorn}},\ }\href
      {\doibase 10.1103/PhysRevLett.132.171002} {\bibfield  {journal} {\bibinfo
      {journal} {Phys. Rev. Lett.}\ }\textbf {\bibinfo {volume} {132}},\ \bibinfo
      {pages} {171002} (\bibinfo {year} {2024})},\ \Eprint
      {http://arxiv.org/abs/2307.02399} {arXiv:2307.02399 [astro-ph.CO]}
      \BibitemShut {NoStop}%
    \bibitem [{\citenamefont {Boyle}\ and\ \citenamefont
      {Steinhardt}(2008)}]{Boyle:2005se}%
      \BibitemOpen
      \bibfield  {author} {\bibinfo {author} {\bibfnamefont {L.~A.}\ \bibnamefont
      {Boyle}}\ and\ \bibinfo {author} {\bibfnamefont {P.~J.}\ \bibnamefont
      {Steinhardt}},\ }\href {\doibase 10.1103/PhysRevD.77.063504} {\bibfield
      {journal} {\bibinfo  {journal} {Phys. Rev. D}\ }\textbf {\bibinfo {volume}
      {77}},\ \bibinfo {pages} {063504} (\bibinfo {year} {2008})},\ \Eprint
      {http://arxiv.org/abs/astro-ph/0512014} {arXiv:astro-ph/0512014} \BibitemShut
      {NoStop}%
    \bibitem [{\citenamefont {Ananda}\ \emph {et~al.}(2007)\citenamefont {Ananda},
      \citenamefont {Clarkson},\ and\ \citenamefont {Wands}}]{Ananda:2006af}%
      \BibitemOpen
      \bibfield  {author} {\bibinfo {author} {\bibfnamefont {K.~N.}\ \bibnamefont
      {Ananda}}, \bibinfo {author} {\bibfnamefont {C.}~\bibnamefont {Clarkson}}, \
      and\ \bibinfo {author} {\bibfnamefont {D.}~\bibnamefont {Wands}},\ }\href
      {\doibase 10.1103/PhysRevD.75.123518} {\bibfield  {journal} {\bibinfo
      {journal} {Phys. Rev. D}\ }\textbf {\bibinfo {volume} {75}},\ \bibinfo
      {pages} {123518} (\bibinfo {year} {2007})},\ \Eprint
      {http://arxiv.org/abs/gr-qc/0612013} {arXiv:gr-qc/0612013} \BibitemShut
      {NoStop}%
    \bibitem [{\citenamefont {Baumann}\ \emph {et~al.}(2007)\citenamefont
      {Baumann}, \citenamefont {Steinhardt}, \citenamefont {Takahashi},\ and\
      \citenamefont {Ichiki}}]{Baumann:2007zm}%
      \BibitemOpen
      \bibfield  {author} {\bibinfo {author} {\bibfnamefont {D.}~\bibnamefont
      {Baumann}}, \bibinfo {author} {\bibfnamefont {P.~J.}\ \bibnamefont
      {Steinhardt}}, \bibinfo {author} {\bibfnamefont {K.}~\bibnamefont
      {Takahashi}}, \ and\ \bibinfo {author} {\bibfnamefont {K.}~\bibnamefont
      {Ichiki}},\ }\href {\doibase 10.1103/PhysRevD.76.084019} {\bibfield
      {journal} {\bibinfo  {journal} {Phys. Rev. D}\ }\textbf {\bibinfo {volume}
      {76}},\ \bibinfo {pages} {084019} (\bibinfo {year} {2007})},\ \Eprint
      {http://arxiv.org/abs/hep-th/0703290} {arXiv:hep-th/0703290} \BibitemShut
      {NoStop}%
    \bibitem [{\citenamefont {Yuan}\ and\ \citenamefont
      {Huang}(2021)}]{Yuan:2021qgz}%
      \BibitemOpen
      \bibfield  {author} {\bibinfo {author} {\bibfnamefont {C.}~\bibnamefont
      {Yuan}}\ and\ \bibinfo {author} {\bibfnamefont {Q.-G.}\ \bibnamefont
      {Huang}},\ }\href {\doibase 10.1016/j.isci.2021.102860} {\bibfield  {journal}
      {\bibinfo  {journal} {iScience}\ }\textbf {\bibinfo {volume} {24}},\ \bibinfo
      {pages} {102860} (\bibinfo {year} {2021})},\ \Eprint
      {http://arxiv.org/abs/2103.04739} {arXiv:2103.04739 [astro-ph.GA]}
      \BibitemShut {NoStop}%
    \bibitem [{\citenamefont {Saito}\ and\ \citenamefont
      {Yokoyama}(2010)}]{Saito:2009jt}%
      \BibitemOpen
      \bibfield  {author} {\bibinfo {author} {\bibfnamefont {R.}~\bibnamefont
      {Saito}}\ and\ \bibinfo {author} {\bibfnamefont {J.}~\bibnamefont
      {Yokoyama}},\ }\href {\doibase 10.1143/PTP.126.351} {\bibfield  {journal}
      {\bibinfo  {journal} {Prog. Theor. Phys.}\ }\textbf {\bibinfo {volume}
      {123}},\ \bibinfo {pages} {867} (\bibinfo {year} {2010})},\ \bibinfo {note}
      {[Erratum: Prog.Theor.Phys. 126, 351--352 (2011)]},\ \Eprint
      {http://arxiv.org/abs/0912.5317} {arXiv:0912.5317 [astro-ph.CO]} \BibitemShut
      {NoStop}%
    \bibitem [{\citenamefont {Wang}\ \emph {et~al.}(2019)\citenamefont {Wang},
      \citenamefont {Terada},\ and\ \citenamefont {Kohri}}]{Wang:2019kaf}%
      \BibitemOpen
      \bibfield  {author} {\bibinfo {author} {\bibfnamefont {S.}~\bibnamefont
      {Wang}}, \bibinfo {author} {\bibfnamefont {T.}~\bibnamefont {Terada}}, \ and\
      \bibinfo {author} {\bibfnamefont {K.}~\bibnamefont {Kohri}},\ }\href
      {\doibase 10.1103/PhysRevD.99.103531} {\bibfield  {journal} {\bibinfo
      {journal} {Phys. Rev. D}\ }\textbf {\bibinfo {volume} {99}},\ \bibinfo
      {pages} {103531} (\bibinfo {year} {2019})},\ \bibinfo {note} {[Erratum:
      Phys.Rev.D 101, 069901 (2020)]},\ \Eprint {http://arxiv.org/abs/1903.05924}
      {arXiv:1903.05924 [astro-ph.CO]} \BibitemShut {NoStop}%
    \bibitem [{\citenamefont {Zhao}\ and\ \citenamefont
      {Wang}(2023)}]{Zhao:2022kvz}%
      \BibitemOpen
      \bibfield  {author} {\bibinfo {author} {\bibfnamefont {Z.-C.}\ \bibnamefont
      {Zhao}}\ and\ \bibinfo {author} {\bibfnamefont {S.}~\bibnamefont {Wang}},\
      }\href {\doibase 10.3390/universe9040157} {\bibfield  {journal} {\bibinfo
      {journal} {Universe}\ }\textbf {\bibinfo {volume} {9}},\ \bibinfo {pages}
      {157} (\bibinfo {year} {2023})},\ \Eprint {http://arxiv.org/abs/2211.09450}
      {arXiv:2211.09450 [astro-ph.CO]} \BibitemShut {NoStop}%
    \bibitem [{\citenamefont {Papanikolaou}\ \emph {et~al.}(2021)\citenamefont
      {Papanikolaou}, \citenamefont {Vennin},\ and\ \citenamefont
      {Langlois}}]{Papanikolaou:2020qtd}%
      \BibitemOpen
      \bibfield  {author} {\bibinfo {author} {\bibfnamefont {T.}~\bibnamefont
      {Papanikolaou}}, \bibinfo {author} {\bibfnamefont {V.}~\bibnamefont
      {Vennin}}, \ and\ \bibinfo {author} {\bibfnamefont {D.}~\bibnamefont
      {Langlois}},\ }\href {\doibase 10.1088/1475-7516/2021/03/053} {\bibfield
      {journal} {\bibinfo  {journal} {JCAP}\ }\textbf {\bibinfo {volume} {03}},\
      \bibinfo {pages} {053} (\bibinfo {year} {2021})},\ \Eprint
      {http://arxiv.org/abs/2010.11573} {arXiv:2010.11573 [astro-ph.CO]}
      \BibitemShut {NoStop}%
    \bibitem [{\citenamefont {Dom\`enech}(2021)}]{Domenech:2021ztg}%
      \BibitemOpen
      \bibfield  {author} {\bibinfo {author} {\bibfnamefont {G.}~\bibnamefont
      {Dom\`enech}},\ }\href {\doibase 10.3390/universe7110398} {\bibfield
      {journal} {\bibinfo  {journal} {Universe}\ }\textbf {\bibinfo {volume} {7}},\
      \bibinfo {pages} {398} (\bibinfo {year} {2021})},\ \Eprint
      {http://arxiv.org/abs/2109.01398} {arXiv:2109.01398 [gr-qc]} \BibitemShut
      {NoStop}%
    \bibitem [{\citenamefont {Cai}\ \emph {et~al.}(2023)\citenamefont {Cai},
      \citenamefont {He}, \citenamefont {Ma}, \citenamefont {Yan},\ and\
      \citenamefont {Yuan}}]{Cai:2023dls}%
      \BibitemOpen
      \bibfield  {author} {\bibinfo {author} {\bibfnamefont {Y.-F.}\ \bibnamefont
      {Cai}}, \bibinfo {author} {\bibfnamefont {X.-C.}\ \bibnamefont {He}},
      \bibinfo {author} {\bibfnamefont {X.-H.}\ \bibnamefont {Ma}}, \bibinfo
      {author} {\bibfnamefont {S.-F.}\ \bibnamefont {Yan}}, \ and\ \bibinfo
      {author} {\bibfnamefont {G.-W.}\ \bibnamefont {Yuan}},\ }\href {\doibase
      10.1016/j.scib.2023.10.027} {\bibfield  {journal} {\bibinfo  {journal} {Sci.
      Bull.}\ }\textbf {\bibinfo {volume} {68}},\ \bibinfo {pages} {2929} (\bibinfo
      {year} {2023})},\ \Eprint {http://arxiv.org/abs/2306.17822} {arXiv:2306.17822
      [gr-qc]} \BibitemShut {NoStop}%
    \bibitem [{\citenamefont {Dom\`enech}\ \emph {et~al.}(2024)\citenamefont
      {Dom\`enech}, \citenamefont {Pi}, \citenamefont {Wang},\ and\ \citenamefont
      {Wang}}]{Domenech:2024rks}%
      \BibitemOpen
      \bibfield  {author} {\bibinfo {author} {\bibfnamefont {G.}~\bibnamefont
      {Dom\`enech}}, \bibinfo {author} {\bibfnamefont {S.}~\bibnamefont {Pi}},
      \bibinfo {author} {\bibfnamefont {A.}~\bibnamefont {Wang}}, \ and\ \bibinfo
      {author} {\bibfnamefont {J.}~\bibnamefont {Wang}},\ }\href {\doibase
      10.1088/1475-7516/2024/08/054} {\bibfield  {journal} {\bibinfo  {journal}
      {JCAP}\ }\textbf {\bibinfo {volume} {08}},\ \bibinfo {pages} {054} (\bibinfo
      {year} {2024})},\ \Eprint {http://arxiv.org/abs/2402.18965} {arXiv:2402.18965
      [astro-ph.CO]} \BibitemShut {NoStop}%
    \bibitem [{\citenamefont {Cecchini}\ \emph {et~al.}(2025)\citenamefont
      {Cecchini}, \citenamefont {Franciolini},\ and\ \citenamefont
      {Pieroni}}]{Cecchini:2025oks}%
      \BibitemOpen
      \bibfield  {author} {\bibinfo {author} {\bibfnamefont {C.}~\bibnamefont
      {Cecchini}}, \bibinfo {author} {\bibfnamefont {G.}~\bibnamefont
      {Franciolini}}, \ and\ \bibinfo {author} {\bibfnamefont {M.}~\bibnamefont
      {Pieroni}},\ }\href@noop {} {\  (\bibinfo {year} {2025})},\ \Eprint
      {http://arxiv.org/abs/2503.10805} {arXiv:2503.10805 [astro-ph.CO]}
      \BibitemShut {NoStop}%
    \bibitem [{\citenamefont {Kosowsky}\ \emph {et~al.}(1992)\citenamefont
      {Kosowsky}, \citenamefont {Turner},\ and\ \citenamefont
      {Watkins}}]{Kosowsky:1992rz}%
      \BibitemOpen
      \bibfield  {author} {\bibinfo {author} {\bibfnamefont {A.}~\bibnamefont
      {Kosowsky}}, \bibinfo {author} {\bibfnamefont {M.~S.}\ \bibnamefont
      {Turner}}, \ and\ \bibinfo {author} {\bibfnamefont {R.}~\bibnamefont
      {Watkins}},\ }\href {\doibase 10.1103/PhysRevLett.69.2026} {\bibfield
      {journal} {\bibinfo  {journal} {Phys. Rev. Lett.}\ }\textbf {\bibinfo
      {volume} {69}},\ \bibinfo {pages} {2026} (\bibinfo {year}
      {1992})}\BibitemShut {NoStop}%
    \bibitem [{\citenamefont {Kamionkowski}\ \emph {et~al.}(1994)\citenamefont
      {Kamionkowski}, \citenamefont {Kosowsky},\ and\ \citenamefont
      {Turner}}]{Kamionkowski:1993fg}%
      \BibitemOpen
      \bibfield  {author} {\bibinfo {author} {\bibfnamefont {M.}~\bibnamefont
      {Kamionkowski}}, \bibinfo {author} {\bibfnamefont {A.}~\bibnamefont
      {Kosowsky}}, \ and\ \bibinfo {author} {\bibfnamefont {M.~S.}\ \bibnamefont
      {Turner}},\ }\href {\doibase 10.1103/PhysRevD.49.2837} {\bibfield  {journal}
      {\bibinfo  {journal} {Phys. Rev. D}\ }\textbf {\bibinfo {volume} {49}},\
      \bibinfo {pages} {2837} (\bibinfo {year} {1994})},\ \Eprint
      {http://arxiv.org/abs/astro-ph/9310044} {arXiv:astro-ph/9310044} \BibitemShut
      {NoStop}%
    \bibitem [{\citenamefont {Caprini}\ \emph {et~al.}(2008)\citenamefont
      {Caprini}, \citenamefont {Durrer},\ and\ \citenamefont
      {Servant}}]{Caprini:2007xq}%
      \BibitemOpen
      \bibfield  {author} {\bibinfo {author} {\bibfnamefont {C.}~\bibnamefont
      {Caprini}}, \bibinfo {author} {\bibfnamefont {R.}~\bibnamefont {Durrer}}, \
      and\ \bibinfo {author} {\bibfnamefont {G.}~\bibnamefont {Servant}},\ }\href
      {\doibase 10.1103/PhysRevD.77.124015} {\bibfield  {journal} {\bibinfo
      {journal} {Phys. Rev. D}\ }\textbf {\bibinfo {volume} {77}},\ \bibinfo
      {pages} {124015} (\bibinfo {year} {2008})},\ \Eprint
      {http://arxiv.org/abs/0711.2593} {arXiv:0711.2593 [astro-ph]} \BibitemShut
      {NoStop}%
    \bibitem [{\citenamefont {Hindmarsh}\ \emph {et~al.}(2014)\citenamefont
      {Hindmarsh}, \citenamefont {Huber}, \citenamefont {Rummukainen},\ and\
      \citenamefont {Weir}}]{Hindmarsh:2013xza}%
      \BibitemOpen
      \bibfield  {author} {\bibinfo {author} {\bibfnamefont {M.}~\bibnamefont
      {Hindmarsh}}, \bibinfo {author} {\bibfnamefont {S.~J.}\ \bibnamefont
      {Huber}}, \bibinfo {author} {\bibfnamefont {K.}~\bibnamefont {Rummukainen}},
      \ and\ \bibinfo {author} {\bibfnamefont {D.~J.}\ \bibnamefont {Weir}},\
      }\href {\doibase 10.1103/PhysRevLett.112.041301} {\bibfield  {journal}
      {\bibinfo  {journal} {Phys. Rev. Lett.}\ }\textbf {\bibinfo {volume} {112}},\
      \bibinfo {pages} {041301} (\bibinfo {year} {2014})},\ \Eprint
      {http://arxiv.org/abs/1304.2433} {arXiv:1304.2433 [hep-ph]} \BibitemShut
      {NoStop}%
    \bibitem [{\citenamefont {Kibble}(1976)}]{Kibble:1976sj}%
      \BibitemOpen
      \bibfield  {author} {\bibinfo {author} {\bibfnamefont {T.~W.~B.}\
      \bibnamefont {Kibble}},\ }\href {\doibase 10.1088/0305-4470/9/8/029}
      {\bibfield  {journal} {\bibinfo  {journal} {J. Phys. A}\ }\textbf {\bibinfo
      {volume} {9}},\ \bibinfo {pages} {1387} (\bibinfo {year} {1976})}\BibitemShut
      {NoStop}%
    \bibitem [{\citenamefont {Vilenkin}(1981{\natexlab{a}})}]{Vilenkin:1981bx}%
      \BibitemOpen
      \bibfield  {author} {\bibinfo {author} {\bibfnamefont {A.}~\bibnamefont
      {Vilenkin}},\ }\href {\doibase 10.1016/0370-2693(81)91144-8} {\bibfield
      {journal} {\bibinfo  {journal} {Phys. Lett. B}\ }\textbf {\bibinfo {volume}
      {107}},\ \bibinfo {pages} {47} (\bibinfo {year}
      {1981}{\natexlab{a}})}\BibitemShut {NoStop}%
    \bibitem [{\citenamefont {Hogan}\ and\ \citenamefont
      {Rees}(1984)}]{Hogan:1984is}%
      \BibitemOpen
      \bibfield  {author} {\bibinfo {author} {\bibfnamefont {C.~J.}\ \bibnamefont
      {Hogan}}\ and\ \bibinfo {author} {\bibfnamefont {M.~J.}\ \bibnamefont
      {Rees}},\ }\href {\doibase 10.1038/311109a0} {\bibfield  {journal} {\bibinfo
      {journal} {Nature}\ }\textbf {\bibinfo {volume} {311}},\ \bibinfo {pages}
      {109} (\bibinfo {year} {1984})}\BibitemShut {NoStop}%
    \bibitem [{\citenamefont {Caldwell}\ and\ \citenamefont
      {Allen}(1992)}]{Caldwell:1991jj}%
      \BibitemOpen
      \bibfield  {author} {\bibinfo {author} {\bibfnamefont {R.~R.}\ \bibnamefont
      {Caldwell}}\ and\ \bibinfo {author} {\bibfnamefont {B.}~\bibnamefont
      {Allen}},\ }\href {\doibase 10.1103/PhysRevD.45.3447} {\bibfield  {journal}
      {\bibinfo  {journal} {Phys. Rev. D}\ }\textbf {\bibinfo {volume} {45}},\
      \bibinfo {pages} {3447} (\bibinfo {year} {1992})}\BibitemShut {NoStop}%
    \bibitem [{\citenamefont {Vilenkin}(1981{\natexlab{b}})}]{Vilenkin:1981zs}%
      \BibitemOpen
      \bibfield  {author} {\bibinfo {author} {\bibfnamefont {A.}~\bibnamefont
      {Vilenkin}},\ }\href {\doibase 10.1103/PhysRevD.23.852} {\bibfield  {journal}
      {\bibinfo  {journal} {Phys. Rev. D}\ }\textbf {\bibinfo {volume} {23}},\
      \bibinfo {pages} {852} (\bibinfo {year} {1981}{\natexlab{b}})}\BibitemShut
      {NoStop}%
    \bibitem [{\citenamefont {Chang}\ \emph {et~al.}(1999)\citenamefont {Chang},
      \citenamefont {Hagmann},\ and\ \citenamefont {Sikivie}}]{Chang:1998tb}%
      \BibitemOpen
      \bibfield  {author} {\bibinfo {author} {\bibfnamefont {S.}~\bibnamefont
      {Chang}}, \bibinfo {author} {\bibfnamefont {C.}~\bibnamefont {Hagmann}}, \
      and\ \bibinfo {author} {\bibfnamefont {P.}~\bibnamefont {Sikivie}},\ }\href
      {\doibase 10.1103/PhysRevD.59.023505} {\bibfield  {journal} {\bibinfo
      {journal} {Phys. Rev. D}\ }\textbf {\bibinfo {volume} {59}},\ \bibinfo
      {pages} {023505} (\bibinfo {year} {1999})},\ \Eprint
      {http://arxiv.org/abs/hep-ph/9807374} {arXiv:hep-ph/9807374} \BibitemShut
      {NoStop}%
    \bibitem [{\citenamefont {Hiramatsu}\ \emph {et~al.}(2010)\citenamefont
      {Hiramatsu}, \citenamefont {Kawasaki},\ and\ \citenamefont
      {Saikawa}}]{Hiramatsu:2010yz}%
      \BibitemOpen
      \bibfield  {author} {\bibinfo {author} {\bibfnamefont {T.}~\bibnamefont
      {Hiramatsu}}, \bibinfo {author} {\bibfnamefont {M.}~\bibnamefont {Kawasaki}},
      \ and\ \bibinfo {author} {\bibfnamefont {K.}~\bibnamefont {Saikawa}},\ }\href
      {\doibase 10.1088/1475-7516/2010/05/032} {\bibfield  {journal} {\bibinfo
      {journal} {JCAP}\ }\textbf {\bibinfo {volume} {05}},\ \bibinfo {pages} {032}
      (\bibinfo {year} {2010})},\ \Eprint {http://arxiv.org/abs/1002.1555}
      {arXiv:1002.1555 [astro-ph.CO]} \BibitemShut {NoStop}%
    \bibitem [{\citenamefont {Chen}\ \emph {et~al.}(2022)\citenamefont {Chen},
      \citenamefont {Wu},\ and\ \citenamefont {Huang}}]{Chen:2022azo}%
      \BibitemOpen
      \bibfield  {author} {\bibinfo {author} {\bibfnamefont {Z.-C.}\ \bibnamefont
      {Chen}}, \bibinfo {author} {\bibfnamefont {Y.-M.}\ \bibnamefont {Wu}}, \ and\
      \bibinfo {author} {\bibfnamefont {Q.-G.}\ \bibnamefont {Huang}},\ }\href
      {\doibase 10.3847/1538-4357/ac86cb} {\bibfield  {journal} {\bibinfo
      {journal} {Astrophys. J.}\ }\textbf {\bibinfo {volume} {936}},\ \bibinfo
      {pages} {20} (\bibinfo {year} {2022})},\ \Eprint
      {http://arxiv.org/abs/2205.07194} {arXiv:2205.07194 [astro-ph.CO]}
      \BibitemShut {NoStop}%
    \bibitem [{\citenamefont {Ashoorioon}\ \emph {et~al.}(2022)\citenamefont
      {Ashoorioon}, \citenamefont {Rezazadeh},\ and\ \citenamefont
      {Rostami}}]{Ashoorioon:2022raz}%
      \BibitemOpen
      \bibfield  {author} {\bibinfo {author} {\bibfnamefont {A.}~\bibnamefont
      {Ashoorioon}}, \bibinfo {author} {\bibfnamefont {K.}~\bibnamefont
      {Rezazadeh}}, \ and\ \bibinfo {author} {\bibfnamefont {A.}~\bibnamefont
      {Rostami}},\ }\href {\doibase 10.1016/j.physletb.2022.137542} {\bibfield
      {journal} {\bibinfo  {journal} {Phys. Lett. B}\ }\textbf {\bibinfo {volume}
      {835}},\ \bibinfo {pages} {137542} (\bibinfo {year} {2022})},\ \Eprint
      {http://arxiv.org/abs/2202.01131} {arXiv:2202.01131 [astro-ph.CO]}
      \BibitemShut {NoStop}%
    \bibitem [{\citenamefont {Bian}\ \emph
      {et~al.}(2024{\natexlab{b}})\citenamefont {Bian}, \citenamefont {Ge},
      \citenamefont {Li}, \citenamefont {Shu},\ and\ \citenamefont
      {Zong}}]{Bian:2022qbh}%
      \BibitemOpen
      \bibfield  {author} {\bibinfo {author} {\bibfnamefont {L.}~\bibnamefont
      {Bian}}, \bibinfo {author} {\bibfnamefont {S.}~\bibnamefont {Ge}}, \bibinfo
      {author} {\bibfnamefont {C.}~\bibnamefont {Li}}, \bibinfo {author}
      {\bibfnamefont {J.}~\bibnamefont {Shu}}, \ and\ \bibinfo {author}
      {\bibfnamefont {J.}~\bibnamefont {Zong}},\ }\href {\doibase
      10.1007/s11433-024-2436-4} {\bibfield  {journal} {\bibinfo  {journal} {Sci.
      China Phys. Mech. Astron.}\ }\textbf {\bibinfo {volume} {67}},\ \bibinfo
      {pages} {110413} (\bibinfo {year} {2024}{\natexlab{b}})},\ \Eprint
      {http://arxiv.org/abs/2212.07871} {arXiv:2212.07871 [hep-ph]} \BibitemShut
      {NoStop}%
    \bibitem [{\citenamefont {Athron}\ \emph {et~al.}(2024)\citenamefont {Athron},
      \citenamefont {Bal\'azs}, \citenamefont {Fowlie}, \citenamefont {Morris},\
      and\ \citenamefont {Wu}}]{Athron:2023xlk}%
      \BibitemOpen
      \bibfield  {author} {\bibinfo {author} {\bibfnamefont {P.}~\bibnamefont
      {Athron}}, \bibinfo {author} {\bibfnamefont {C.}~\bibnamefont {Bal\'azs}},
      \bibinfo {author} {\bibfnamefont {A.}~\bibnamefont {Fowlie}}, \bibinfo
      {author} {\bibfnamefont {L.}~\bibnamefont {Morris}}, \ and\ \bibinfo {author}
      {\bibfnamefont {L.}~\bibnamefont {Wu}},\ }\href {\doibase
      10.1016/j.ppnp.2023.104094} {\bibfield  {journal} {\bibinfo  {journal} {Prog.
      Part. Nucl. Phys.}\ }\textbf {\bibinfo {volume} {135}},\ \bibinfo {pages}
      {104094} (\bibinfo {year} {2024})},\ \Eprint
      {http://arxiv.org/abs/2305.02357} {arXiv:2305.02357 [hep-ph]} \BibitemShut
      {NoStop}%
    \bibitem [{\citenamefont {He}\ \emph {et~al.}(2025)\citenamefont {He},
      \citenamefont {Li}, \citenamefont {Wang},\ and\ \citenamefont
      {Wang}}]{He:2023ado}%
      \BibitemOpen
      \bibfield  {author} {\bibinfo {author} {\bibfnamefont {S.}~\bibnamefont
      {He}}, \bibinfo {author} {\bibfnamefont {L.}~\bibnamefont {Li}}, \bibinfo
      {author} {\bibfnamefont {S.}~\bibnamefont {Wang}}, \ and\ \bibinfo {author}
      {\bibfnamefont {S.-J.}\ \bibnamefont {Wang}},\ }\href {\doibase
      10.1007/s11433-024-2468-x} {\bibfield  {journal} {\bibinfo  {journal} {Sci.
      China Phys. Mech. Astron.}\ }\textbf {\bibinfo {volume} {68}},\ \bibinfo
      {pages} {210411} (\bibinfo {year} {2025})},\ \Eprint
      {http://arxiv.org/abs/2308.07257} {arXiv:2308.07257 [hep-ph]} \BibitemShut
      {NoStop}%
    \bibitem [{\citenamefont {Bertone}\ \emph {et~al.}(2005)\citenamefont
      {Bertone}, \citenamefont {Hooper},\ and\ \citenamefont
      {Silk}}]{Bertone:2004pz}%
      \BibitemOpen
      \bibfield  {author} {\bibinfo {author} {\bibfnamefont {G.}~\bibnamefont
      {Bertone}}, \bibinfo {author} {\bibfnamefont {D.}~\bibnamefont {Hooper}}, \
      and\ \bibinfo {author} {\bibfnamefont {J.}~\bibnamefont {Silk}},\ }\href
      {\doibase 10.1016/j.physrep.2004.08.031} {\bibfield  {journal} {\bibinfo
      {journal} {Phys. Rept.}\ }\textbf {\bibinfo {volume} {405}},\ \bibinfo
      {pages} {279} (\bibinfo {year} {2005})},\ \Eprint
      {http://arxiv.org/abs/hep-ph/0404175} {arXiv:hep-ph/0404175} \BibitemShut
      {NoStop}%
    \bibitem [{\citenamefont {Clowe}\ \emph {et~al.}(2006)\citenamefont {Clowe},
      \citenamefont {Bradac}, \citenamefont {Gonzalez}, \citenamefont {Markevitch},
      \citenamefont {Randall}, \citenamefont {Jones},\ and\ \citenamefont
      {Zaritsky}}]{Clowe:2006eq}%
      \BibitemOpen
      \bibfield  {author} {\bibinfo {author} {\bibfnamefont {D.}~\bibnamefont
      {Clowe}}, \bibinfo {author} {\bibfnamefont {M.}~\bibnamefont {Bradac}},
      \bibinfo {author} {\bibfnamefont {A.~H.}\ \bibnamefont {Gonzalez}}, \bibinfo
      {author} {\bibfnamefont {M.}~\bibnamefont {Markevitch}}, \bibinfo {author}
      {\bibfnamefont {S.~W.}\ \bibnamefont {Randall}}, \bibinfo {author}
      {\bibfnamefont {C.}~\bibnamefont {Jones}}, \ and\ \bibinfo {author}
      {\bibfnamefont {D.}~\bibnamefont {Zaritsky}},\ }\href {\doibase
      10.1086/508162} {\bibfield  {journal} {\bibinfo  {journal} {Astrophys. J.
      Lett.}\ }\textbf {\bibinfo {volume} {648}},\ \bibinfo {pages} {L109}
      (\bibinfo {year} {2006})},\ \Eprint {http://arxiv.org/abs/astro-ph/0608407}
      {arXiv:astro-ph/0608407} \BibitemShut {NoStop}%
    \bibitem [{\citenamefont {Ullio}\ \emph {et~al.}(2001)\citenamefont {Ullio},
      \citenamefont {Zhao},\ and\ \citenamefont {Kamionkowski}}]{Ullio:2001fb}%
      \BibitemOpen
      \bibfield  {author} {\bibinfo {author} {\bibfnamefont {P.}~\bibnamefont
      {Ullio}}, \bibinfo {author} {\bibfnamefont {H.}~\bibnamefont {Zhao}}, \ and\
      \bibinfo {author} {\bibfnamefont {M.}~\bibnamefont {Kamionkowski}},\ }\href
      {\doibase 10.1103/PhysRevD.64.043504} {\bibfield  {journal} {\bibinfo
      {journal} {Phys. Rev. D}\ }\textbf {\bibinfo {volume} {64}},\ \bibinfo
      {pages} {043504} (\bibinfo {year} {2001})},\ \Eprint
      {http://arxiv.org/abs/astro-ph/0101481} {arXiv:astro-ph/0101481} \BibitemShut
      {NoStop}%
    \bibitem [{\citenamefont {Yuan}\ \emph {et~al.}(2022)\citenamefont {Yuan},
      \citenamefont {Chen}, \citenamefont {Shen}, \citenamefont {Guo},
      \citenamefont {Ding}, \citenamefont {Huang},\ and\ \citenamefont
      {Yuan}}]{Yuan:2021mzi}%
      \BibitemOpen
      \bibfield  {author} {\bibinfo {author} {\bibfnamefont {G.-W.}\ \bibnamefont
      {Yuan}}, \bibinfo {author} {\bibfnamefont {Z.-F.}\ \bibnamefont {Chen}},
      \bibinfo {author} {\bibfnamefont {Z.-Q.}\ \bibnamefont {Shen}}, \bibinfo
      {author} {\bibfnamefont {W.-Q.}\ \bibnamefont {Guo}}, \bibinfo {author}
      {\bibfnamefont {R.}~\bibnamefont {Ding}}, \bibinfo {author} {\bibfnamefont
      {X.}~\bibnamefont {Huang}}, \ and\ \bibinfo {author} {\bibfnamefont
      {Q.}~\bibnamefont {Yuan}},\ }\href {\doibase 10.1007/JHEP04(2022)018}
      {\bibfield  {journal} {\bibinfo  {journal} {JHEP}\ }\textbf {\bibinfo
      {volume} {04}},\ \bibinfo {pages} {018} (\bibinfo {year} {2022})},\ \Eprint
      {http://arxiv.org/abs/2106.05901} {arXiv:2106.05901 [hep-ph]} \BibitemShut
      {NoStop}%
    \bibitem [{\citenamefont {Shen}\ \emph {et~al.}(2023)\citenamefont {Shen},
      \citenamefont {Yuan}, \citenamefont {Jiang}, \citenamefont {Tsai},
      \citenamefont {Yuan},\ and\ \citenamefont {Fan}}]{Shen:2023kkm}%
      \BibitemOpen
      \bibfield  {author} {\bibinfo {author} {\bibfnamefont {Z.-Q.}\ \bibnamefont
      {Shen}}, \bibinfo {author} {\bibfnamefont {G.-W.}\ \bibnamefont {Yuan}},
      \bibinfo {author} {\bibfnamefont {C.-Z.}\ \bibnamefont {Jiang}}, \bibinfo
      {author} {\bibfnamefont {Y.-L.~S.}\ \bibnamefont {Tsai}}, \bibinfo {author}
      {\bibfnamefont {Q.}~\bibnamefont {Yuan}}, \ and\ \bibinfo {author}
      {\bibfnamefont {Y.-Z.}\ \bibnamefont {Fan}},\ }\href {\doibase
      10.1093/mnras/stad3282} {\bibfield  {journal} {\bibinfo  {journal} {Mon. Not.
      Roy. Astron. Soc.}\ }\textbf {\bibinfo {volume} {527}},\ \bibinfo {pages}
      {3196} (\bibinfo {year} {2023})},\ \Eprint {http://arxiv.org/abs/2303.09284}
      {arXiv:2303.09284 [astro-ph.GA]} \BibitemShut {NoStop}%
    \bibitem [{\citenamefont {Aghaie}\ \emph {et~al.}(2024)\citenamefont {Aghaie},
      \citenamefont {Armando}, \citenamefont {Dondarini},\ and\ \citenamefont
      {Panci}}]{Aghaie:2023lan}%
      \BibitemOpen
      \bibfield  {author} {\bibinfo {author} {\bibfnamefont {M.}~\bibnamefont
      {Aghaie}}, \bibinfo {author} {\bibfnamefont {G.}~\bibnamefont {Armando}},
      \bibinfo {author} {\bibfnamefont {A.}~\bibnamefont {Dondarini}}, \ and\
      \bibinfo {author} {\bibfnamefont {P.}~\bibnamefont {Panci}},\ }\href
      {\doibase 10.1103/PhysRevD.109.103030} {\bibfield  {journal} {\bibinfo
      {journal} {Phys. Rev. D}\ }\textbf {\bibinfo {volume} {109}},\ \bibinfo
      {pages} {103030} (\bibinfo {year} {2024})},\ \Eprint
      {http://arxiv.org/abs/2308.04590} {arXiv:2308.04590 [astro-ph.CO]}
      \BibitemShut {NoStop}%
    \bibitem [{\citenamefont {Hu}\ \emph {et~al.}(2025)\citenamefont {Hu},
      \citenamefont {Cai},\ and\ \citenamefont {Wang}}]{Hu:2023oiu}%
      \BibitemOpen
      \bibfield  {author} {\bibinfo {author} {\bibfnamefont {L.}~\bibnamefont
      {Hu}}, \bibinfo {author} {\bibfnamefont {R.-G.}\ \bibnamefont {Cai}}, \ and\
      \bibinfo {author} {\bibfnamefont {S.-J.}\ \bibnamefont {Wang}},\ }\href
      {\doibase 10.1088/1475-7516/2025/02/067} {\bibfield  {journal} {\bibinfo
      {journal} {JCAP}\ }\textbf {\bibinfo {volume} {02}},\ \bibinfo {pages} {067}
      (\bibinfo {year} {2025})},\ \Eprint {http://arxiv.org/abs/2312.14041}
      {arXiv:2312.14041 [gr-qc]} \BibitemShut {NoStop}%
    \bibitem [{\citenamefont {Dosopoulou}(2024)}]{Dosopoulou:2023umg}%
      \BibitemOpen
      \bibfield  {author} {\bibinfo {author} {\bibfnamefont {F.}~\bibnamefont
      {Dosopoulou}},\ }\href {\doibase 10.1103/PhysRevD.110.083027} {\bibfield
      {journal} {\bibinfo  {journal} {Phys. Rev. D}\ }\textbf {\bibinfo {volume}
      {110}},\ \bibinfo {pages} {083027} (\bibinfo {year} {2024})},\ \Eprint
      {http://arxiv.org/abs/2305.17281} {arXiv:2305.17281 [astro-ph.HE]}
      \BibitemShut {NoStop}%
    \bibitem [{\citenamefont {Zhang}\ \emph {et~al.}(2025)\citenamefont {Zhang},
      \citenamefont {Yuan},\ and\ \citenamefont {Tang}}]{Zhang:2025mdl}%
      \BibitemOpen
      \bibfield  {author} {\bibinfo {author} {\bibfnamefont {Z.-C.}\ \bibnamefont
      {Zhang}}, \bibinfo {author} {\bibfnamefont {H.-C.}\ \bibnamefont {Yuan}}, \
      and\ \bibinfo {author} {\bibfnamefont {Y.}~\bibnamefont {Tang}},\ }\href@noop
      {} {\  (\bibinfo {year} {2025})},\ \Eprint {http://arxiv.org/abs/2503.02573}
      {arXiv:2503.02573 [astro-ph.GA]} \BibitemShut {NoStop}%
    \bibitem [{\citenamefont {Mitra}\ \emph {et~al.}(2025)\citenamefont {Mitra},
      \citenamefont {Speeney}, \citenamefont {Chakraborty},\ and\ \citenamefont
      {Berti}}]{Mitra:2025tag}%
      \BibitemOpen
      \bibfield  {author} {\bibinfo {author} {\bibfnamefont {S.}~\bibnamefont
      {Mitra}}, \bibinfo {author} {\bibfnamefont {N.}~\bibnamefont {Speeney}},
      \bibinfo {author} {\bibfnamefont {S.}~\bibnamefont {Chakraborty}}, \ and\
      \bibinfo {author} {\bibfnamefont {E.}~\bibnamefont {Berti}},\ }\href@noop {}
      {\  (\bibinfo {year} {2025})},\ \Eprint {http://arxiv.org/abs/2505.04697}
      {arXiv:2505.04697 [gr-qc]} \BibitemShut {NoStop}%
    \bibitem [{\citenamefont {Eda}\ \emph {et~al.}(2015)\citenamefont {Eda},
      \citenamefont {Itoh}, \citenamefont {Kuroyanagi},\ and\ \citenamefont
      {Silk}}]{Eda:2014kra}%
      \BibitemOpen
      \bibfield  {author} {\bibinfo {author} {\bibfnamefont {K.}~\bibnamefont
      {Eda}}, \bibinfo {author} {\bibfnamefont {Y.}~\bibnamefont {Itoh}}, \bibinfo
      {author} {\bibfnamefont {S.}~\bibnamefont {Kuroyanagi}}, \ and\ \bibinfo
      {author} {\bibfnamefont {J.}~\bibnamefont {Silk}},\ }\href {\doibase
      10.1103/PhysRevD.91.044045} {\bibfield  {journal} {\bibinfo  {journal} {Phys.
      Rev. D}\ }\textbf {\bibinfo {volume} {91}},\ \bibinfo {pages} {044045}
      (\bibinfo {year} {2015})},\ \Eprint {http://arxiv.org/abs/1408.3534}
      {arXiv:1408.3534 [gr-qc]} \BibitemShut {NoStop}%
    \bibitem [{\citenamefont {Yue}\ and\ \citenamefont {Cao}(2019)}]{Yue:2019ozq}%
      \BibitemOpen
      \bibfield  {author} {\bibinfo {author} {\bibfnamefont {X.-J.}\ \bibnamefont
      {Yue}}\ and\ \bibinfo {author} {\bibfnamefont {Z.}~\bibnamefont {Cao}},\
      }\href {\doibase 10.1103/PhysRevD.100.043013} {\bibfield  {journal} {\bibinfo
       {journal} {Phys. Rev. D}\ }\textbf {\bibinfo {volume} {100}},\ \bibinfo
      {pages} {043013} (\bibinfo {year} {2019})},\ \Eprint
      {http://arxiv.org/abs/1908.10241} {arXiv:1908.10241 [astro-ph.HE]}
      \BibitemShut {NoStop}%
    \bibitem [{\citenamefont {Li}\ \emph {et~al.}(2022)\citenamefont {Li},
      \citenamefont {Tang},\ and\ \citenamefont {Wu}}]{Li:2021pxf}%
      \BibitemOpen
      \bibfield  {author} {\bibinfo {author} {\bibfnamefont {G.-L.}\ \bibnamefont
      {Li}}, \bibinfo {author} {\bibfnamefont {Y.}~\bibnamefont {Tang}}, \ and\
      \bibinfo {author} {\bibfnamefont {Y.-L.}\ \bibnamefont {Wu}},\ }\href
      {\doibase 10.1007/s11433-022-1930-9} {\bibfield  {journal} {\bibinfo
      {journal} {Sci. China Phys. Mech. Astron.}\ }\textbf {\bibinfo {volume}
      {65}},\ \bibinfo {pages} {100412} (\bibinfo {year} {2022})},\ \Eprint
      {http://arxiv.org/abs/2112.14041} {arXiv:2112.14041 [astro-ph.CO]}
      \BibitemShut {NoStop}%
    \bibitem [{\citenamefont {Zhang}\ and\ \citenamefont
      {Tang}(2024)}]{Zhang:2024hrq}%
      \BibitemOpen
      \bibfield  {author} {\bibinfo {author} {\bibfnamefont {Z.-C.}\ \bibnamefont
      {Zhang}}\ and\ \bibinfo {author} {\bibfnamefont {Y.}~\bibnamefont {Tang}},\
      }\href {\doibase 10.1103/PhysRevD.110.103008} {\bibfield  {journal} {\bibinfo
       {journal} {Phys. Rev. D}\ }\textbf {\bibinfo {volume} {110}},\ \bibinfo
      {pages} {103008} (\bibinfo {year} {2024})},\ \Eprint
      {http://arxiv.org/abs/2403.18529} {arXiv:2403.18529 [astro-ph.GA]}
      \BibitemShut {NoStop}%
    \bibitem [{\citenamefont {Cheng}\ \emph {et~al.}(2025)\citenamefont {Cheng},
      \citenamefont {Cao},\ and\ \citenamefont {Tang}}]{Cheng:2024mgl}%
      \BibitemOpen
      \bibfield  {author} {\bibinfo {author} {\bibfnamefont {Y.-Z.}\ \bibnamefont
      {Cheng}}, \bibinfo {author} {\bibfnamefont {Y.}~\bibnamefont {Cao}}, \ and\
      \bibinfo {author} {\bibfnamefont {Y.}~\bibnamefont {Tang}},\ }\href {\doibase
      10.1103/PhysRevD.111.083010} {\bibfield  {journal} {\bibinfo  {journal}
      {Phys. Rev. D}\ }\textbf {\bibinfo {volume} {111}},\ \bibinfo {pages}
      {083010} (\bibinfo {year} {2025})},\ \Eprint
      {http://arxiv.org/abs/2411.03095} {arXiv:2411.03095 [gr-qc]} \BibitemShut
      {NoStop}%
    \bibitem [{\citenamefont {Daniel}\ \emph {et~al.}(2025)\citenamefont {Daniel},
      \citenamefont {Pardo},\ and\ \citenamefont {Sagunski}}]{Daniel:2025mna}%
      \BibitemOpen
      \bibfield  {author} {\bibinfo {author} {\bibfnamefont {M.}~\bibnamefont
      {Daniel}}, \bibinfo {author} {\bibfnamefont {K.}~\bibnamefont {Pardo}}, \
      and\ \bibinfo {author} {\bibfnamefont {L.}~\bibnamefont {Sagunski}},\
      }\href@noop {} {\  (\bibinfo {year} {2025})},\ \Eprint
      {http://arxiv.org/abs/2501.13601} {arXiv:2501.13601 [astro-ph.HE]}
      \BibitemShut {NoStop}%
    \bibitem [{\citenamefont {Xie}\ and\ \citenamefont {Tang}(2025)}]{Xie:2025udx}%
      \BibitemOpen
      \bibfield  {author} {\bibinfo {author} {\bibfnamefont {Z.-M.}\ \bibnamefont
      {Xie}}\ and\ \bibinfo {author} {\bibfnamefont {Y.}~\bibnamefont {Tang}},\
      }\href@noop {} {\  (\bibinfo {year} {2025})},\ \Eprint
      {http://arxiv.org/abs/2501.12574} {arXiv:2501.12574 [gr-qc]} \BibitemShut
      {NoStop}%
    \bibitem [{\citenamefont {Feng}\ \emph {et~al.}(2025)\citenamefont {Feng},
      \citenamefont {Tang},\ and\ \citenamefont {Wu}}]{Feng:2025fkc}%
      \BibitemOpen
      \bibfield  {author} {\bibinfo {author} {\bibfnamefont {C.}~\bibnamefont
      {Feng}}, \bibinfo {author} {\bibfnamefont {Y.}~\bibnamefont {Tang}}, \ and\
      \bibinfo {author} {\bibfnamefont {Y.-L.}\ \bibnamefont {Wu}},\ }\href@noop {}
      {\  (\bibinfo {year} {2025})},\ \Eprint {http://arxiv.org/abs/2506.02937}
      {arXiv:2506.02937 [astro-ph.GA]} \BibitemShut {NoStop}%
    \bibitem [{\citenamefont {Alonso-\'Alvarez}\ \emph {et~al.}(2024)\citenamefont
      {Alonso-\'Alvarez}, \citenamefont {Cline},\ and\ \citenamefont
      {Dewar}}]{Alonso-Alvarez:2024gdz}%
      \BibitemOpen
      \bibfield  {author} {\bibinfo {author} {\bibfnamefont {G.}~\bibnamefont
      {Alonso-\'Alvarez}}, \bibinfo {author} {\bibfnamefont {J.~M.}\ \bibnamefont
      {Cline}}, \ and\ \bibinfo {author} {\bibfnamefont {C.}~\bibnamefont
      {Dewar}},\ }\href {\doibase 10.1103/PhysRevLett.133.021401} {\bibfield
      {journal} {\bibinfo  {journal} {Phys. Rev. Lett.}\ }\textbf {\bibinfo
      {volume} {133}},\ \bibinfo {pages} {021401} (\bibinfo {year} {2024})},\
      \Eprint {http://arxiv.org/abs/2401.14450} {arXiv:2401.14450 [astro-ph.CO]}
      \BibitemShut {NoStop}%
    \bibitem [{\citenamefont {Chen}\ and\ \citenamefont
      {Tang}(2025)}]{Chen:2025jch}%
      \BibitemOpen
      \bibfield  {author} {\bibinfo {author} {\bibfnamefont {M.-C.}\ \bibnamefont
      {Chen}}\ and\ \bibinfo {author} {\bibfnamefont {Y.}~\bibnamefont {Tang}},\
      }\href@noop {} {\  (\bibinfo {year} {2025})},\ \Eprint
      {http://arxiv.org/abs/2505.09219} {arXiv:2505.09219 [astro-ph.GA]}
      \BibitemShut {NoStop}%
    \bibitem [{\citenamefont {{Chan}}\ and\ \citenamefont
      {{Lee}}(2024)}]{2024ApJ...962L..40C}%
      \BibitemOpen
      \bibfield  {author} {\bibinfo {author} {\bibfnamefont {M.~H.}\ \bibnamefont
      {{Chan}}}\ and\ \bibinfo {author} {\bibfnamefont {C.~M.}\ \bibnamefont
      {{Lee}}},\ }\href {\doibase 10.3847/2041-8213/ad2465} {\bibfield  {journal}
      {\bibinfo  {journal} {\apjl}\ }\textbf {\bibinfo {volume} {962}},\ \bibinfo
      {eid} {L40} (\bibinfo {year} {2024})},\ \Eprint
      {http://arxiv.org/abs/2402.03751} {arXiv:2402.03751 [astro-ph.GA]}
      \BibitemShut {NoStop}%
    \bibitem [{\citenamefont {Fischer}\ and\ \citenamefont
      {Sagunski}(2024)}]{Fischer:2024dte}%
      \BibitemOpen
      \bibfield  {author} {\bibinfo {author} {\bibfnamefont {M.~S.}\ \bibnamefont
      {Fischer}}\ and\ \bibinfo {author} {\bibfnamefont {L.}~\bibnamefont
      {Sagunski}},\ }\href {\doibase 10.1051/0004-6361/202451304} {\bibfield
      {journal} {\bibinfo  {journal} {Astron. Astrophys.}\ }\textbf {\bibinfo
      {volume} {690}},\ \bibinfo {pages} {A299} (\bibinfo {year} {2024})},\ \Eprint
      {http://arxiv.org/abs/2405.19392} {arXiv:2405.19392 [astro-ph.CO]}
      \BibitemShut {NoStop}%
    \bibitem [{\citenamefont {Sesana}\ \emph {et~al.}(2006)\citenamefont {Sesana},
      \citenamefont {Haardt},\ and\ \citenamefont {Madau}}]{Sesana:2006xw}%
      \BibitemOpen
      \bibfield  {author} {\bibinfo {author} {\bibfnamefont {A.}~\bibnamefont
      {Sesana}}, \bibinfo {author} {\bibfnamefont {F.}~\bibnamefont {Haardt}}, \
      and\ \bibinfo {author} {\bibfnamefont {P.}~\bibnamefont {Madau}},\ }\href
      {\doibase 10.1086/507596} {\bibfield  {journal} {\bibinfo  {journal}
      {Astrophys. J.}\ }\textbf {\bibinfo {volume} {651}},\ \bibinfo {pages} {392}
      (\bibinfo {year} {2006})},\ \Eprint {http://arxiv.org/abs/astro-ph/0604299}
      {arXiv:astro-ph/0604299} \BibitemShut {NoStop}%
    \bibitem [{\citenamefont {Sesana}\ \emph {et~al.}(2007)\citenamefont {Sesana},
      \citenamefont {Haardt},\ and\ \citenamefont {Madau}}]{Sesana:2006ne}%
      \BibitemOpen
      \bibfield  {author} {\bibinfo {author} {\bibfnamefont {A.}~\bibnamefont
      {Sesana}}, \bibinfo {author} {\bibfnamefont {F.}~\bibnamefont {Haardt}}, \
      and\ \bibinfo {author} {\bibfnamefont {P.}~\bibnamefont {Madau}},\ }\href
      {\doibase 10.1086/513016} {\bibfield  {journal} {\bibinfo  {journal}
      {Astrophys. J.}\ }\textbf {\bibinfo {volume} {660}},\ \bibinfo {pages} {546}
      (\bibinfo {year} {2007})},\ \Eprint {http://arxiv.org/abs/astro-ph/0612265}
      {arXiv:astro-ph/0612265} \BibitemShut {NoStop}%
    \bibitem [{\citenamefont {Sesana}\ \emph
      {et~al.}(2008{\natexlab{b}})\citenamefont {Sesana}, \citenamefont {Haardt},\
      and\ \citenamefont {Madau}}]{Sesana:2007vr}%
      \BibitemOpen
      \bibfield  {author} {\bibinfo {author} {\bibfnamefont {A.}~\bibnamefont
      {Sesana}}, \bibinfo {author} {\bibfnamefont {F.}~\bibnamefont {Haardt}}, \
      and\ \bibinfo {author} {\bibfnamefont {P.}~\bibnamefont {Madau}},\ }\href
      {\doibase 10.1086/590651} {\bibfield  {journal} {\bibinfo  {journal}
      {Astrophys. J.}\ }\textbf {\bibinfo {volume} {686}},\ \bibinfo {pages} {432}
      (\bibinfo {year} {2008}{\natexlab{b}})},\ \Eprint
      {http://arxiv.org/abs/0710.4301} {arXiv:0710.4301 [astro-ph]} \BibitemShut
      {NoStop}%
    \bibitem [{\citenamefont {Sesana}\ \emph {et~al.}(2009)\citenamefont {Sesana},
      \citenamefont {Vecchio},\ and\ \citenamefont {Volonteri}}]{Sesana:2008xk}%
      \BibitemOpen
      \bibfield  {author} {\bibinfo {author} {\bibfnamefont {A.}~\bibnamefont
      {Sesana}}, \bibinfo {author} {\bibfnamefont {A.}~\bibnamefont {Vecchio}}, \
      and\ \bibinfo {author} {\bibfnamefont {M.}~\bibnamefont {Volonteri}},\ }\href
      {\doibase 10.1111/j.1365-2966.2009.14499.x} {\bibfield  {journal} {\bibinfo
      {journal} {Mon. Not. Roy. Astron. Soc.}\ }\textbf {\bibinfo {volume} {394}},\
      \bibinfo {pages} {2255} (\bibinfo {year} {2009})},\ \Eprint
      {http://arxiv.org/abs/0809.3412} {arXiv:0809.3412 [astro-ph]} \BibitemShut
      {NoStop}%
    \bibitem [{\citenamefont {Shannon}\ \emph {et~al.}(2015)\citenamefont {Shannon}
      \emph {et~al.}}]{Shannon:2015ect}%
      \BibitemOpen
      \bibfield  {author} {\bibinfo {author} {\bibfnamefont {R.~M.}\ \bibnamefont
      {Shannon}} \emph {et~al.},\ }\href {\doibase 10.1126/science.aab1910}
      {\bibfield  {journal} {\bibinfo  {journal} {Science}\ }\textbf {\bibinfo
      {volume} {349}},\ \bibinfo {pages} {1522} (\bibinfo {year} {2015})},\ \Eprint
      {http://arxiv.org/abs/1509.07320} {arXiv:1509.07320 [astro-ph.CO]}
      \BibitemShut {NoStop}%
    \bibitem [{\citenamefont {Kelley}\ \emph {et~al.}(2017)\citenamefont {Kelley},
      \citenamefont {Blecha},\ and\ \citenamefont {Hernquist}}]{Kelley:2016gse}%
      \BibitemOpen
      \bibfield  {author} {\bibinfo {author} {\bibfnamefont {L.~Z.}\ \bibnamefont
      {Kelley}}, \bibinfo {author} {\bibfnamefont {L.}~\bibnamefont {Blecha}}, \
      and\ \bibinfo {author} {\bibfnamefont {L.}~\bibnamefont {Hernquist}},\ }\href
      {\doibase 10.1093/mnras/stw2452} {\bibfield  {journal} {\bibinfo  {journal}
      {Mon. Not. Roy. Astron. Soc.}\ }\textbf {\bibinfo {volume} {464}},\ \bibinfo
      {pages} {3131} (\bibinfo {year} {2017})},\ \Eprint
      {http://arxiv.org/abs/1606.01900} {arXiv:1606.01900 [astro-ph.HE]}
      \BibitemShut {NoStop}%
    \bibitem [{\citenamefont {Chandrasekhar}(1943)}]{1943ApJ....97..255C}%
      \BibitemOpen
      \bibfield  {author} {\bibinfo {author} {\bibfnamefont {S.}~\bibnamefont
      {Chandrasekhar}},\ }\href {\doibase 10.1086/144517} {\bibfield  {journal}
      {\bibinfo  {journal} {Astrophys. J.}\ }\textbf {\bibinfo {volume} {97}},\
      \bibinfo {pages} {255} (\bibinfo {year} {1943})}\BibitemShut {NoStop}%
    \bibitem [{\citenamefont {Navarro}\ \emph {et~al.}(1997)\citenamefont
      {Navarro}, \citenamefont {Frenk},\ and\ \citenamefont
      {White}}]{Navarro:1996gj}%
      \BibitemOpen
      \bibfield  {author} {\bibinfo {author} {\bibfnamefont {J.~F.}\ \bibnamefont
      {Navarro}}, \bibinfo {author} {\bibfnamefont {C.~S.}\ \bibnamefont {Frenk}},
      \ and\ \bibinfo {author} {\bibfnamefont {S.~D.~M.}\ \bibnamefont {White}},\
      }\href {\doibase 10.1086/304888} {\bibfield  {journal} {\bibinfo  {journal}
      {Astrophys. J.}\ }\textbf {\bibinfo {volume} {490}},\ \bibinfo {pages} {493}
      (\bibinfo {year} {1997})},\ \Eprint {http://arxiv.org/abs/astro-ph/9611107}
      {arXiv:astro-ph/9611107} \BibitemShut {NoStop}%
    \bibitem [{\citenamefont {Aghanim}\ \emph {et~al.}(2020)\citenamefont {Aghanim}
      \emph {et~al.}}]{Planck:2018vyg}%
      \BibitemOpen
      \bibfield  {author} {\bibinfo {author} {\bibfnamefont {N.}~\bibnamefont
      {Aghanim}} \emph {et~al.} (\bibinfo {collaboration} {Planck}),\ }\href
      {\doibase 10.1051/0004-6361/201833910} {\bibfield  {journal} {\bibinfo
      {journal} {Astron. Astrophys.}\ }\textbf {\bibinfo {volume} {641}},\ \bibinfo
      {pages} {A6} (\bibinfo {year} {2020})},\ \bibinfo {note} {[Erratum:
      Astron.Astrophys. 652, C4 (2021)]},\ \Eprint
      {http://arxiv.org/abs/1807.06209} {arXiv:1807.06209 [astro-ph.CO]}
      \BibitemShut {NoStop}%
    \bibitem [{\citenamefont {Klypin}\ \emph {et~al.}(2016)\citenamefont {Klypin},
      \citenamefont {Yepes}, \citenamefont {Gottlober}, \citenamefont {Prada},\
      and\ \citenamefont {Hess}}]{Klypin:2014kpa}%
      \BibitemOpen
      \bibfield  {author} {\bibinfo {author} {\bibfnamefont {A.}~\bibnamefont
      {Klypin}}, \bibinfo {author} {\bibfnamefont {G.}~\bibnamefont {Yepes}},
      \bibinfo {author} {\bibfnamefont {S.}~\bibnamefont {Gottlober}}, \bibinfo
      {author} {\bibfnamefont {F.}~\bibnamefont {Prada}}, \ and\ \bibinfo {author}
      {\bibfnamefont {S.}~\bibnamefont {Hess}},\ }\href {\doibase
      10.1093/mnras/stw248} {\bibfield  {journal} {\bibinfo  {journal} {Mon. Not.
      Roy. Astron. Soc.}\ }\textbf {\bibinfo {volume} {457}},\ \bibinfo {pages}
      {4340} (\bibinfo {year} {2016})},\ \Eprint {http://arxiv.org/abs/1411.4001}
      {arXiv:1411.4001 [astro-ph.CO]} \BibitemShut {NoStop}%
    \bibitem [{\citenamefont {Girelli}\ \emph {et~al.}(2020)\citenamefont
      {Girelli}, \citenamefont {Pozzetti}, \citenamefont {Bolzonella},
      \citenamefont {Giocoli}, \citenamefont {Marulli},\ and\ \citenamefont
      {Baldi}}]{Girelli:2020goz}%
      \BibitemOpen
      \bibfield  {author} {\bibinfo {author} {\bibfnamefont {G.}~\bibnamefont
      {Girelli}}, \bibinfo {author} {\bibfnamefont {L.}~\bibnamefont {Pozzetti}},
      \bibinfo {author} {\bibfnamefont {M.}~\bibnamefont {Bolzonella}}, \bibinfo
      {author} {\bibfnamefont {C.}~\bibnamefont {Giocoli}}, \bibinfo {author}
      {\bibfnamefont {F.}~\bibnamefont {Marulli}}, \ and\ \bibinfo {author}
      {\bibfnamefont {M.}~\bibnamefont {Baldi}},\ }\href {\doibase
      10.1051/0004-6361/201936329} {\bibfield  {journal} {\bibinfo  {journal}
      {Astron. Astrophys.}\ }\textbf {\bibinfo {volume} {634}},\ \bibinfo {pages}
      {A135} (\bibinfo {year} {2020})},\ \Eprint {http://arxiv.org/abs/2001.02230}
      {arXiv:2001.02230 [astro-ph.CO]} \BibitemShut {NoStop}%
    \bibitem [{\citenamefont {Chen}\ \emph {et~al.}(2019)\citenamefont {Chen},
      \citenamefont {Sesana},\ and\ \citenamefont {Conselice}}]{Chen:2018znx}%
      \BibitemOpen
      \bibfield  {author} {\bibinfo {author} {\bibfnamefont {S.}~\bibnamefont
      {Chen}}, \bibinfo {author} {\bibfnamefont {A.}~\bibnamefont {Sesana}}, \ and\
      \bibinfo {author} {\bibfnamefont {C.~J.}\ \bibnamefont {Conselice}},\ }\href
      {\doibase 10.1093/mnras/stz1722} {\bibfield  {journal} {\bibinfo  {journal}
      {Mon. Not. Roy. Astron. Soc.}\ }\textbf {\bibinfo {volume} {488}},\ \bibinfo
      {pages} {401} (\bibinfo {year} {2019})},\ \Eprint
      {http://arxiv.org/abs/1810.04184} {arXiv:1810.04184 [astro-ph.GA]}
      \BibitemShut {NoStop}%
    \bibitem [{\citenamefont {Kormendy}\ and\ \citenamefont
      {Ho}(2013)}]{Kormendy:2013dxa}%
      \BibitemOpen
      \bibfield  {author} {\bibinfo {author} {\bibfnamefont {J.}~\bibnamefont
      {Kormendy}}\ and\ \bibinfo {author} {\bibfnamefont {L.~C.}\ \bibnamefont
      {Ho}},\ }\href {\doibase 10.1146/annurev-astro-082708-101811} {\bibfield
      {journal} {\bibinfo  {journal} {Ann. Rev. Astron. Astrophys.}\ }\textbf
      {\bibinfo {volume} {51}},\ \bibinfo {pages} {511} (\bibinfo {year} {2013})},\
      \Eprint {http://arxiv.org/abs/1304.7762} {arXiv:1304.7762 [astro-ph.CO]}
      \BibitemShut {NoStop}%
    \bibitem [{\citenamefont {Lacroix}(2018)}]{Lacroix:2018zmg}%
      \BibitemOpen
      \bibfield  {author} {\bibinfo {author} {\bibfnamefont {T.}~\bibnamefont
      {Lacroix}},\ }\href {\doibase 10.1051/0004-6361/201832652} {\bibfield
      {journal} {\bibinfo  {journal} {Astron. Astrophys.}\ }\textbf {\bibinfo
      {volume} {619}},\ \bibinfo {pages} {A46} (\bibinfo {year} {2018})},\ \Eprint
      {http://arxiv.org/abs/1801.01308} {arXiv:1801.01308 [astro-ph.GA]}
      \BibitemShut {NoStop}%
    \bibitem [{\citenamefont {Gondolo}\ and\ \citenamefont
      {Silk}(1999)}]{Gondolo:1999ef}%
      \BibitemOpen
      \bibfield  {author} {\bibinfo {author} {\bibfnamefont {P.}~\bibnamefont
      {Gondolo}}\ and\ \bibinfo {author} {\bibfnamefont {J.}~\bibnamefont {Silk}},\
      }\href {\doibase 10.1103/PhysRevLett.83.1719} {\bibfield  {journal} {\bibinfo
       {journal} {Phys. Rev. Lett.}\ }\textbf {\bibinfo {volume} {83}},\ \bibinfo
      {pages} {1719} (\bibinfo {year} {1999})},\ \Eprint
      {http://arxiv.org/abs/astro-ph/9906391} {arXiv:astro-ph/9906391} \BibitemShut
      {NoStop}%
    \bibitem [{\citenamefont {Phinney}(2001)}]{Phinney:2001di}%
      \BibitemOpen
      \bibfield  {author} {\bibinfo {author} {\bibfnamefont {E.~S.}\ \bibnamefont
      {Phinney}},\ }\href@noop {} {\  (\bibinfo {year} {2001})},\ \Eprint
      {http://arxiv.org/abs/astro-ph/0108028} {arXiv:astro-ph/0108028} \BibitemShut
      {NoStop}%
    \bibitem [{\citenamefont {Chen}\ \emph
      {et~al.}(2017{\natexlab{a}})\citenamefont {Chen}, \citenamefont {Middleton},
      \citenamefont {Sesana}, \citenamefont {Del~Pozzo},\ and\ \citenamefont
      {Vecchio}}]{Chen:2016kax}%
      \BibitemOpen
      \bibfield  {author} {\bibinfo {author} {\bibfnamefont {S.}~\bibnamefont
      {Chen}}, \bibinfo {author} {\bibfnamefont {H.}~\bibnamefont {Middleton}},
      \bibinfo {author} {\bibfnamefont {A.}~\bibnamefont {Sesana}}, \bibinfo
      {author} {\bibfnamefont {W.}~\bibnamefont {Del~Pozzo}}, \ and\ \bibinfo
      {author} {\bibfnamefont {A.}~\bibnamefont {Vecchio}},\ }\href {\doibase
      10.1093/mnras/stx475} {\bibfield  {journal} {\bibinfo  {journal} {Mon. Not.
      Roy. Astron. Soc.}\ }\textbf {\bibinfo {volume} {468}},\ \bibinfo {pages}
      {404} (\bibinfo {year} {2017}{\natexlab{a}})},\ \bibinfo {note} {[Erratum:
      Mon.Not.Roy.Astron.Soc. 469, 2455--2456 (2017)]},\ \Eprint
      {http://arxiv.org/abs/1612.02826} {arXiv:1612.02826 [astro-ph.HE]}
      \BibitemShut {NoStop}%
    \bibitem [{\citenamefont {Chen}\ \emph
      {et~al.}(2017{\natexlab{b}})\citenamefont {Chen}, \citenamefont {Sesana},\
      and\ \citenamefont {Del~Pozzo}}]{Chen:2016zyo}%
      \BibitemOpen
      \bibfield  {author} {\bibinfo {author} {\bibfnamefont {S.}~\bibnamefont
      {Chen}}, \bibinfo {author} {\bibfnamefont {A.}~\bibnamefont {Sesana}}, \ and\
      \bibinfo {author} {\bibfnamefont {W.}~\bibnamefont {Del~Pozzo}},\ }\href
      {\doibase 10.1093/mnras/stx1093} {\bibfield  {journal} {\bibinfo  {journal}
      {Mon. Not. Roy. Astron. Soc.}\ }\textbf {\bibinfo {volume} {470}},\ \bibinfo
      {pages} {1738} (\bibinfo {year} {2017}{\natexlab{b}})},\ \Eprint
      {http://arxiv.org/abs/1612.00455} {arXiv:1612.00455 [astro-ph.CO]}
      \BibitemShut {NoStop}%
    \bibitem [{\citenamefont {{Foreman-Mackey}}\ \emph {et~al.}(2013)\citenamefont
      {{Foreman-Mackey}}, \citenamefont {{Hogg}}, \citenamefont {{Lang}},\ and\
      \citenamefont {{Goodman}}}]{emcee}%
      \BibitemOpen
      \bibfield  {author} {\bibinfo {author} {\bibfnamefont {D.}~\bibnamefont
      {{Foreman-Mackey}}}, \bibinfo {author} {\bibfnamefont {D.~W.}\ \bibnamefont
      {{Hogg}}}, \bibinfo {author} {\bibfnamefont {D.}~\bibnamefont {{Lang}}}, \
      and\ \bibinfo {author} {\bibfnamefont {J.}~\bibnamefont {{Goodman}}},\ }\href
      {\doibase 10.1086/670067} {\bibfield  {journal} {\bibinfo  {journal} {PASP}\
      }\textbf {\bibinfo {volume} {125}},\ \bibinfo {pages} {306} (\bibinfo {year}
      {2013})},\ \Eprint {http://arxiv.org/abs/1202.3665} {1202.3665} \BibitemShut
      {NoStop}%
    \end{thebibliography}
%

\end{document}